\documentclass[nofootinbib,notitlepage,superscriptaddress,10pt,aps,pra,twocolumn]{revtex4-1}\usepackage{graphicx}
\usepackage{amsmath,amssymb,wasysym}
\usepackage{color}
\usepackage{ulem}
\pdfoutput=1
\graphicspath{{figures/}}

\def\Msun{m$_{\odot} \,$}

\newcommand{\apjl}{Ap. J. Lett. }
\newcommand{\aj}{A. J. }

\newcommand{\physrep}{Phys. Rep.}
\newcommand{\mnras}{Mon. not. RAS.}

\newcommand{\aap}{Astron. \& Astrophys.}

\begin{document}
\title{Implications of GRB 130603B and its macronova for r-process nucleosynthesis}
\author{ Tsvi Piran$^1$}
\author{Oleg Korobkin$^2$}
\author{Stephan Rosswog$^2$} 
\affiliation{$^{1}$Racah Institute of Physics, The Hebrew University, Jerusalem 91904, Israel\\
$^2${Department of Astronomy and Oskar Klein Centre}, Stockholm University, AlbaNova, SE-10691 Stockholm, Sweden}

\date{\today}

\begin{abstract}

The tentative identification of a Li-Paczynski macronova following the short GRB 130603B indicated that
a  few hundredths of a solar mass of neutron star matter were  ejected and that this ejected mass
has  radioactively decayed { into  heavy r-process elements}. If correct, this  confirms long
standing predictions \citep{eichler89} that on the one hand,  sGRBs are produced in compact binary mergers  (CBMs)
and on the other hand that these events are significant and possibly dominant sources of the heavy 
($A>130$) r-process { nuclei}. Assuming that this interpretation is correct we obtain a lower limit of 
$0.02 m_\odot$ on the ejected mass. Using the current estimates of the rate of sGRBs and with a  beaming 
factor of 50, mergers associated with sGRBs can produce all the observed heavy r-process material 
in the Universe.  We confront this conclusion with  cosmochemistry and show that even though such 
events are rare, mixing is sufficient to account for the current homogeneous distribution of 
r-process material in the Galaxy. However, the appearance of significant  amounts of Eu in some very 
low metallicity stars requires that some mergers took place very early on, namely with a very short 
time delay after the earliest star formation episodes.  Alternatively, an additional early r-process
source may have contributed  at that early stage.  Finally, we note that evidence for short 
lived $^{244}$Pu in the very early solar system suggests that a merger
of this kind took place within the vicinity of the solar system shortly (a few hundred million years) 
before its formation.
\end{abstract}

\maketitle

\section{Introduction}

Rapid neutron capture  (``r-process'') has been known as a basic formation process
for the heaviest elements in the cosmos since the seminal work of BBFH \cite{burbidge57} 
and Cameron \cite{cameron57}. The question in which astrophysical environment it 
actually occurs, however, has puzzled astrophysics ever since. Traditionally,
core-collapse supernovae  have been considered as the 
favored production site \citep{takahashi94,woosley94,hoffman97,freiburghaus99a,farouqi10},
see \cite{arcones13a} for a recent review. 
As an alternative site, Lattimer and Schramm \cite{lattimer74} suggested the decompression of cold 
nuclear matter ejected during the tidal disruption of a neutron star
by a stellar mass black hole, a process that occurs in a similar way
during the merger of two neutron stars. 

Eichler et al. \cite{eichler89} placed compact binary mergers (CBMs)
in a broader astrophysical context.  They suggested that in addition 
to bursts of gravitational waves and neutrinos CBMs  are 
the engines of a subclass of Gamma-ray Bursts (GRBs).  Estimating the rate of 
mergers  they also suggested that these events could be a major source
of r-process material.  
In the first detailed calculation of mass ejection from a merger, 
Rosswog et al. \cite{rosswog99} found that $\sim 10^{-2}$ \Msun become unbound during a merger 
event due to gravitational torques and hydrodynamic interaction (this is 
referred to as ``dynamic ejecta'' in the following).  In their follow-up
work  \cite{freiburghaus99b}  they 
performed the first network calculations based 
hydrodynamics simulations which showed that the resulting abundance patterns 
agree well with the observed solar system abundances for $A > 130$ and, folding the ejecta masses with the estimated merger rates, 
this indicated  that, indeed, CBMs could represent a major source of 
cosmic r-process.
At the same time Li and Paczynski \cite{LP98} suggested that  the radioactive decay of  the neutron-rich
nuclei in these ``dynamic ejecta'' would produce a macronova (also 
referred to as kilonova): a short lived optical - IR weak  supernova-like signal.
The suggested triple link between GRBs, heavy element nucleosynthesis  and  macronovae will 
be the backbone of this paper.

During the past decade much effort has been invested in understanding
the relevance and the implications of CBM r-process.
\citep{LP98,rosswog00,ruffert01,rosswog05a,oechslin07a,surman08,metzger08,beloborodov08, lee09, roberts11,goriely11a,
korobkin12a,metzger12a, caballero12,wanajo12,malkus12,kasen13a,barnes13a,rosswog13a,rosswog13b, bauswein13a,kyutoku13,surman13a}. 
The picture
that emerges from this wealth of studies is the following: a) the dynamic ejecta 
with their extreme neutron-richness provide excellent conditions for heavy ($A>130$) 
r-process and the resulting abundance pattern is largely independent
of the specifics of the merging binary system \citep[e.g.][]{korobkin12a}] and b) CBMs
provide additional nucleosynthesis channels such as neutrino-driven winds 
and the final dissolution of the accretion disk which likely complement the 
nucleosynthesis from the dynamic ejecta \citep[e.g.][]{fernandez13}.

On the other hand, essentially all recent studies agree that the originally
favored scenario, core-collapse supernovae, is seriously challenged, at the
very least for the production of elements heavier than $A=110$ 
\citep{arcones07,roberts10,fischer10,huedepohl10,roberts12a,martinez_pinedo12a, wanajo13a,
arcones13a} (An exception may possibly be very rapidly rotating,
highly magnetized progenitor stars that can also produce favorable conditions
for r-process \citep{winteler12a}.).
Despite the growing consistency among different theory/simulation results 
-- in favour of CBMs, against core-collapse supernovae --
some reservations against the neutron star merger scenario have remained, 
mainly due to the unsettled question whether or not they are consistent 
with the chemical evolution of galaxies \citep{qian00,argast04,matteucci13}.

Studies of macronova evolution 
\citep{LP98,kulkarni05,metzger10a,
piran13a,barnes13a,tanaka13a,grossman13a} 
revealed its prospects for detectability. Most notable is the recent realization 
\citep{kasen13a}
that the opacity of the ejecta will be fairly large as it will be dominated by Lanthanides.
This has lead to a qualitative shift in the expected light curve. 
While earlier estimates, based on iron group opacities, predicted a UV-optical
signal at around half a day \citep{LP98,kulkarni05,metzger10a,
piran13a}, recent ones used opacities of Lanthanides and obtained a
weaker IR signal peaking at a week \citep{barnes13a,tanaka13a,grossman13a}.

On June 3rd 2013 the {\it Swift} satellite detected a relatively nearby  
sGRB \citep{Melandri13} 
at a redshift of 0.356. With a duration of $0.18 \pm 0.02$s 
and with a hard spectrum GRB 130603B is a genuine non-collapsar \citep{Bromberg13}. 
It had a regular optical  and X-ray afterglow with some   X-ray excess at $>1$ day 
\citep{fong13b}, perhaps indicating a fallback accretion \cite{rosswog07a}. Later radio afterglow 
observations suggested a jet break \citep{fong13b}. At 9 days  after the burst
($\approx 6.6$ days  in the local rest frame)  HST  \citep{Tanvir+13,Berger+13} 
detected a nIR point source with an apparent magnitude of 
$H_{\rm 160,AB} =25.73 \pm 0.2$ ({$M_{\rm J,AB}\approx-15.35$}), corresponding 
to  an intrinsic luminosity of $\approx 10^{41}$erg/sec. The upper limit on the R 
band emission at the same time, $R_{\rm 606,AB} > 28.5$, suggests that the 
regular afterglow has decayed by this time. The  IR excess at 9 days 
after the burst  was interpreted by both groups as
tentative evidence for a Li-Paczynski macronova. 
In the following we consider the implications of this interpretation for the r-process 
nucleosynthesis.

If this interpretation is correct it has several interesting implications. 
It provides the first {\it direct}\footnote{So far there is only indirect 
evidence that short GRBs arise from mergers: nature of hosts, position 
within the hosts and overall rate, \citep{Nakar07, Fong+13,guetta06,NGF06}.} 
evidence that 
sGRBs arise from CBMs. Furthermore, within the 
macronova model the IR emission arises from the combination of the heating of
the neutron star debris via radioactive decays \citep{LP98} and the large 
opacity dominated by Lanthanides \citep{kasen13a}. Thus, 
the observations indicate that (i) CBMs indeed power sGRBs 
and (ii) they are also the sources of a substantial fraction of the heaviest
r-process nuclei.  This  provides the impetus to revisit the role of compact 
binary mergers for the formation of the heaviest elements in the Universe. 

In the following  we derive limits on the amount of ejected material in \S 2. In \S 3 we discuss  the implications for cosmochemistry, in \S 4 the implication to the composition  of the early solar system  and in \S 5 those for the CBMs.

\section{Limits on the ejected mass}

The nIR luminosity and the late time heating rate  \citep{korobkin12}
 yield a lower limit on the ejected mass: 
\begin{equation} 
m_{\rm ej} >  0.02  (\epsilon/0.5)^{-1} m_\odot ,
\label{Mejc}
\end{equation}
where $\epsilon$ is the fraction of radioactive energy in electrons or positrons and $\gamma$-rays.
We assume conservatively 
that this fraction is converted locally to heat  and radiated  as nIR. 
While conservative this strict lower limit  is in  rough agreement with various estimates of the macronova 
light curve \citep{barnes13a,tanaka13a,grossman13a}. 
It is subject only to uncertainties in the estimates of the radioactive 
heating rate and of the  fraction of energy that is captured and re-emitted. 
At the maximum  only about half of the mass contributes to the emission,  therefore
we use in the following  0.04 \Msun as an estimate for the ejected mass. 

The radiation escape condition is $m_{ej} \kappa_L / (4 \pi (v_{\rm ej} t)^2) \approx c/v $, where $\kappa_L$ is the opacity 
and $v$ is the typical velocity of the ejected material. This 
yields  $v_{\rm ej}=0.2 c~ (m_{\rm ej}/0.04 m_\odot)$ and  a size of  
$3 \cdot 10^{15} (m_{\rm ej}/0.04 m_\odot)$~cm.  The corresponding black body temperature 
$\sim 1500^o$K with a peak emission at $\sim2 \mu$M, is consistent with  the observed nIR peak. 
The peak time 
would have been at approximately 
$4.5 (m_{\rm ej}/0.04 m_\odot)^{1/2} (\kappa_L/10 {\rm gm/cm}^2)^{1/2} (v/0.2c)^{-1/2}$ days,
suggesting a lucky coincidence in the choice of the observing epoch.

\section{r-process Nucleosynthesis}

 Assuming solar-system abundances \citep{kappler89} everywhere
the Milky Way contains $M^{A>130} = 1.6 \cdot 10^4 \; m_\odot $ 
in ``heavy" ($A > 130$) r-process material.
We assume that a)
all mergers eject a comparable amount of r-process material to the one observed in this 
event 
and b) only CBMs are responsible for the heavy r-process. With  $m_{\rm ej}= 0.04 m_\odot$  
\begin{equation} 
N  =  4  \cdot 10^5 \left( \frac{M^{A>130}} {1.6 \cdot 10^4 m_\odot}\right) \left(\frac{ 0.04 m_\odot }{ m_{\rm ej}}\right)
\label{Nmerger}
\end{equation}
mergers should have taken place in the Galaxy. 
The corresponding merger rate,
\begin{eqnarray} 
R_{\rm CBM} =  
300 \left(\frac{  0.04 m_\odot}  {m_{\rm ej}} \right) \left( \frac{M^{A>130}} { 1.6 \cdot 10^4 m_\odot}\right) {\rm Gpc}^{-3}  {\rm yr}^{-1},&
\label{Rnuc}
\end{eqnarray}
should be compared with  the sGRB rate  
\cite{Wanderman13}    $R_{\rm SGRB} = 6 \pm 2\;{\rm Gpc}^{-3} {\rm yr}^{-1}$ which agrees within a factor of 2 with various previous estimates \citep{guetta06,NGF06,GS09}
(which is inversely proportional to a lower sGRB
luminosity limit, here taken as $5 \cdot  10^{49}$~erg/s, 
roughly the lower value of luminosity for all sGRBs with known redshifts).
This implies that only 1 out of 50 mergers produces an observable
sGRB{, which}
could arise
either from some mergers failing to produce
a detectable burst 
or from a finite beaming angle. 
The latter is consistent with the estimated  beaming correction of 
$\sim 100$ for this  burst \cite{fong13b}, but this number is not very well constrained. 
Estimates based on the observed binary pulsars in the
Galaxy are highly uncertain, with values in the  range 
$R_{\rm pulsars} = 20 - 20 000\;{\rm Gpc}^{-3} {\rm yr}^{-1} $ 
\citep{NPS91,Phinney91,KalEtal04,KalEtal04a,
abadie10}
and a canonical value of $\approx 800\;{\rm Gpc}^{-3} {\rm yr}^{-1} $. 
These
values are consistent with the above rates.

Turning now to  cosmochemical evolution we note that there exists an observed 
correlation between abundances of r-process elements and Fe at 
${\rm [Fe/H]} >  -2.5$ \citep{gratton94,crawford98,mcwilliam95a}.
Since Fe enrichment is controlled by type Ia SNe occurring at a 
much higher rate, widely varying degrees of mixing of the r-process ejecta in
an already existing  merger remnant with Fe produced by fresh SNe would result in
large scatter in the r-process abundances over a broad range of [Fe/H].

By comparison with supernovae Qian \cite{qian00} argued that the dynamic ejecta 
from a single event cannot  mix  with more than 
$m_{\rm mix}\approx{3\cdot10^4 m_\odot}$
material.
If the ejecta are all r-process this would lead to a fraction of 
$3 \cdot 10^{-7}  (m_{\rm ej}/0.01 m_\odot) (3 \cdot 10^4 m_\odot/m_{\rm mix}) $
which, as he argues, would strongly disfavor neutron star mergers
since this fraction is much larger than observed. Furthermore, as one expects only about $5 \cdot 10^5$ mergers in the Galaxy and there are
about $10^7$ ``cells" of $3\cdot10^4 m_\odot$ not all material 
would be mixed with  r-process products. Argast et al. \cite{argast04} carried out a
detailed simulation and reached the same conclusion on the basis of
similar ideas.

We argue here that substantial mixing could resolve this problem. This 
could arise simply due to  galactic rotation, which leads to significant mixing 
on a rotational time scale of $\sim 200$~Myr, due to Rosetta-type  or 
volume-filling orbits of stars/debris, not uncommon in a typical galaxy, and 
due to turbulent diffusion. For the latter, with a typical turbulent velocity $v_{\rm turb} \approx c_s
\approx 10^6$~cm/s one expects  significant mixing on a $\sim 200$~Myr 
time scale. This depends critically on the size of the turbulent cells. 
We have conservatively assumed a size of 1~pc. Diffusion will be much more
effective, like the square root of this quantity, if these cells are
 larger. At early times galaxies were more turbulent and \citep{cresci09} observe
turbulent motion of up to 150~km/sec in milky way like galaxies at $z\sim 2$, with 
a corresponding diffusion length of a kpc.  This could easily lead to significant mixing
within one or two rotations. 
Finally,  if the neutron stars in compact binary systems
receive, like single neutron stars, a kick at birth, they may travel a few 
kpc out of the midplane  \citep[see e.g.][]{narayan92,fryer99a,bloom02,rosswog03c},
consistent with the observation of the locations of sGRBs \citep{fong10}.
This would mean that the ejecta are sprayed over a larger volume and can
easily mix with a substantially larger amount of mass.

Thus  a uniform distribution of r-process material can be
established. The observations of a large scatter in [Eu/Fe] abundances
at low metallically stars are  consistent with this picture in which  heavy
r-process material is produced in rare events and in large amounts.
At early times some material was exposed to such events while other 
material was not.

Abundances of r-process elements have been observed in halo and disk
stars covering a metallicity range ${\rm [Fe/H]} \approx  -3.1-0.5$   
\citep{woolf95,Shetrone96,sneden00,Burris00,Cayrel01, Hill02}.  
This requires a significant fraction of r-process nucleosynthesis to take place within a few Myr. 
Some population synthesis models  \citep{belczynski02b} suggest such a rapid population of mergers.
However, this requirement is in a direct contrast with estimates of the time delay between sGRBs 
and the SFR \citep[e.g.][]{Wanderman13}. Typical delays are found 
by all groups to be of order of 3 Gyr. The distribution around this time delay is rather narrow, not 
larger than 1.5 Gyr.

Can one resolve this inconsistency? The in-spiral time is very sensitive
to the initial orbital period and eccentricity.
For typical ns$^2$ parameters a modest variation by a factor of 7 in the initial separation (from
$2 \cdot 10^{11}$~cm to $3 \cdot 10^{10}$~cm would change the merger time
all the way from 3 Gyr to 1 Myr, which is practically instantaneous. Even the
larger separations, that correspond to long merger times of several Gyr are
significantly smaller than the two original stellar radii. This implies that
the system must have undergone a common envelope phase and the initial separation
of the binary (and hence the merger time
delay) depends critically on this  poorly understood  phase. 
Most observed sGRBs are at relatively low redshifts. Conditions could 
have been different at earlier epochs, where low metallicity may have
lead to shorter delays. Note that such low metallicity systems could have arisen without 
contamination by the SNe that formed the neutrons stars as 
some neutron stars (e.g. J0737-3039B) formed in a collapse with practically no mass ejection
\cite{piran05a}.
Alternatively it  is possible that a second channel of special rare type of  SNe \citep{winteler12a}, 
that can operate only at very early times, produces the earliest $A>130$  material. 
However, as these SNe require very fast rotation and hence can operate only in low 
metallicity stars, they are unlikely to produce {the bulk} of the r-process material.

\section{Implications to the early Solar System}

{ Evidence for $^{244}$Pu, with a half life time
$t_{1/2}= 81$~Myr ($\tau$= 117 Myr) in the Early Solar System \cite{wasserburg06}  
provides further
independent evidence for the rarity of formation episodes of r-process
material.} This shows 
actinide production in the vicinity (in time and space) of the Solar
System prior to its formation. However, searches \cite{paul01,walner14} for  
traces of  $^{244}$Pu isotope in deep-sea crust and sediment accumulated
over the last $\sim25$~Myr have found none.  The current limits on deep-sea 
crust and sediments are that the $^{244}$Pu abundance is lower by a factor 
30-200 than expected from a uniform production model, assuming actinide 
production in SNe and reasonable assumptions on interstellar ISM deposition. 
These  findings suggest that: (a) formation of $^{244}$Pu and heavy
r-process in general occured in rare episodes such as CBMs,
and (b) one of CBMs took place in the vicinity of our early Solar
system less than a few hundred million years just prior to its formation.

\section{Implication to compact binary mergers}
\label{sec:binary}

The derived lower limit on $m_{\rm ej}$ is consistent with the models of
\cite{barnes13a} and \cite{tanaka13a}. But at least if we take all current
numbers at their face values, this would make a double neutron star merger as
somewhat unlikely ``engine"  for this sGRB and the subsequent macronova emission. Different
groups find the following numbers { from numerical simulations}:
$m_{\rm ej} < 0.04 \; m_\odot $, \citep{rosswog13b}; 
$m_{\rm ej} < 0.02 \; m_\odot $, \citep{bauswein13a}; 
$m_{\rm ej} < 0.02 \; m_\odot $ \citep{hotokezaka13b}. 
The lower limit that we find here is just at the upper end
of those estimates, and  it ignores additional mass loss due to the nuclear
energy release. 
Mergers of neutron star black hole systems, in contrast, could comfortably
eject such large amounts of mass 
\citep{rosswog05a,lovelace13,kyutoku13,deaton13,hotokezaka13b}.
Dynamic collisions between black holes and neutron stars that may occur in 
globular clusters were also found to eject masses in the required range 
\citep{lee10a,rosswog13a}. { The involvement of a black hole
would also be consistent with the large estimated ejecta velocity around 
$0.2 c$, which naturally occurs in both mergers and collisions 
involving a black hole, while a typical nsns merger only leads to $v_{\rm ej} \approx 0.1 c$,
see Tab.1 in \cite{rosswog13b}.
However, this interpretation of bhns merger is in some tension} with the fact 
that the needed rates, as well  as the estimated sGRB rates agree roughly 
with the observation-based estimates of Galactic ns$^2$.

\section{Conclusions}

The tentative identification  of a macronova following a sGRBs has numerous 
far-reaching implications. It provides a direct proof that sGRBs arise from compact 
binary mergers, as  \cite{eichler89} suggested long ago. The late IR signal clearly 
indicates  high-$A$ material and demonstrates  that  these events are associated 
with  the production of r-process material. The assertion that  they are indeed the 
sources of most of the heavy r-process material has additional implications. It  
narrows down both the mass ejection from  these sources as well as their rate and 
cosmic evolution. 
(i) The lower limit on implied mass $m_{\rm ej}>   0.02 m_\odot$, which 
 is only subject to errors in the estimated radioactive heating 
rate  is 
at the very high end of ns$^2$ mergers \citep[see e.g.][]{piran13a,rosswog13b}. 
Together with the large estimated ejecta velocity, this points to the involvement of
a black hole, either in a merger or in a dynamical collision \citep{rosswog13a}.
(ii) The required rate is consistent with sGRB rate estimates with a reasonable 
 beaming factor of order 50.   This is consistent with current estimates of beaming of sGRBs
\citep{fong13b}.

Favorable indications towards this idea include: (a) uniformity in abundances, suggesting 
a regular source - freely expanding neutron star material is natural (b) problems with SNe 
{ nucleosynthesis} (c) large fluctuation at early times - indicating a rare process that produces 
a significant { amount} of material at each episode. In the past arguments against compact 
binary mergers as sources of r-process material have been mostly based on indications 
from chemical evolution of the Galaxy. Observations suggest that the distribution of heavy 
r-process material becomes rather uniform at a rather early stage. They also indicate early 
deposition of such material.  The observed uniformity can be obtained if mixing within the 
galaxy is rapid enough.  This is plausibly mediated by turbulent mixing, 
differential rotation, and the fast motion of some binaries within the Galaxy. One 
has to recall that the early galaxy was much more erratic than the Milky Way we observe today 
and that turbulent velocities of 150 km/sec have been observed in $z\sim 2$ galaxies. The 
various mechanisms could easily mix the galaxy over a few rotational periods. 

The CBM r-process model requires rapid mergers took place at a very early stage. Such a population of rapid merger has been suggested by population 
synthesis \citep{belczynski02b}. However, the observed redshift distribution and luminosities 
of sGRBs suggest a typical spiraling-in time of 2-3 Gyr. A possible way to reconcile the two 
observations is that an earlier stellar population (e.g. of low metallicity) had somewhat lower
initial separations. The required change in the initial separation is  only moderate.  
Alternatively it is possible that the earliest r-process material is produced via another channel \citep{winteler12a}.

It is interesting to note that evidence for $^{244}$Pu in the Early Solar System \cite{wasserburg06}
and with lack of evidence for deposition of this isotope from the ISM during the last 25 Myr \cite{paul01,walner14}
suggest that this isotope, as well as other heavy r-process elements 
are produced in rare episodes. If heavy r-process material is indeed produced in CMBs this implies that 
such an event  took place in  the vicinity of the solar system just prior to its formation.

To conclude we note that the idea that 
the recently observed GRB130603B with a macronova was accompanied by a significant mass 
ejection could possibly be tested if the surrounding density is not too low (current estimates are 
$\approx5\cdot10^{-3}-30\;{\rm cm}^{-3}$,  \cite{fong13b}.). In such a case we 
expect a significant (a few tens of $\mu Jy$) radio signal to arise from the interaction of this 
ejecta with the circum-merger matter \citep{NP11,piran13a}. This signal would rise on a time 
scale of months or a year and should be detectable by the Extended Very Large Array (EVLA). A detection would be a 
direct confirmation. However, even if for some reason (e.g. low circum-merger density) this 
signal { should not be detectable} we can expect a detection of another  sGRB at a comparable 
distance within a year or two. This scenario has a clear prediction and the identification of 
the accompanying IR macronova is expected.

TP thanks M. Paul for helpful discussions. 
The research was  supported by an ERC advanced grant (GRBs) and by the  I-CORE 
(grant No 1829/12) [T.P.],
by DFG grant RO-3399,  AOBJ-584282 and by the Swedish 
Research Council (VR)  grant 621-2012-4870 [
S.R. and O.K.] and by Compstar [S.R.].


\begin{thebibliography}{95}%
\makeatletter
\providecommand \@ifxundefined [1]{%
 \@ifx{#1\undefined}
}%
\providecommand \@ifnum [1]{%
 \ifnum #1\expandafter \@firstoftwo
 \else \expandafter \@secondoftwo
 \fi
}%
\providecommand \@ifx [1]{%
 \ifx #1\expandafter \@firstoftwo
 \else \expandafter \@secondoftwo
 \fi
}%
\providecommand \natexlab [1]{#1}%
\providecommand \enquote  [1]{``#1''}%
\providecommand \bibnamefont  [1]{#1}%
\providecommand \bibfnamefont [1]{#1}%
\providecommand \citenamefont [1]{#1}%
\providecommand \href@noop [0]{\@secondoftwo}%
\providecommand \href [0]{\begingroup \@sanitize@url \@href}%
\providecommand \@href[1]{\@@startlink{#1}\@@href}%
\providecommand \@@href[1]{\endgroup#1\@@endlink}%
\providecommand \@sanitize@url [0]{\catcode `\\12\catcode `\$12\catcode
  `\&12\catcode `\#12\catcode `\^12\catcode `\_12\catcode `\%12\relax}%
\providecommand \@@startlink[1]{}%
\providecommand \@@endlink[0]{}%
\providecommand \url  [0]{\begingroup\@sanitize@url \@url }%
\providecommand \@url [1]{\endgroup\@href {#1}{\urlprefix }}%
\providecommand \urlprefix  [0]{URL }%
\providecommand \Eprint [0]{\href }%
\providecommand \doibase [0]{http://dx.doi.org/}%
\providecommand \selectlanguage [0]{\@gobble}%
\providecommand \bibinfo  [0]{\@secondoftwo}%
\providecommand \bibfield  [0]{\@secondoftwo}%
\providecommand \translation [1]{[#1]}%
\providecommand \BibitemOpen [0]{}%
\providecommand \bibitemStop [0]{}%
\providecommand \bibitemNoStop [0]{.\EOS\space}%
\providecommand \EOS [0]{\spacefactor3000\relax}%
\providecommand \BibitemShut  [1]{\csname bibitem#1\endcsname}%
\let\auto@bib@innerbib\@empty
\bibitem [{\citenamefont {{Eichler}}\ \emph {et~al.}(1989)\citenamefont
  {{Eichler}}, \citenamefont {{Livio}}, \citenamefont {{Piran}},\ and\
  \citenamefont {{Schramm}}}]{eichler89}%
  \BibitemOpen
  \bibfield  {author} {\bibinfo {author} {\bibfnamefont {D.}~\bibnamefont
  {{Eichler}}}, \bibinfo {author} {\bibfnamefont {M.}~\bibnamefont {{Livio}}},
  \bibinfo {author} {\bibfnamefont {T.}~\bibnamefont {{Piran}}}, \ and\
  \bibinfo {author} {\bibfnamefont {D.~N.}\ \bibnamefont {{Schramm}}},\ }\href
  {\doibase 10.1038/340126a0} {\bibfield  {journal} {\bibinfo  {journal}
  {\nat}\ }\textbf {\bibinfo {volume} {340}},\ \bibinfo {pages} {126} (\bibinfo
  {year} {1989})}%
\bibitem [{\citenamefont {Burbidge}\ \emph {et~al.}(1957)\citenamefont
  {Burbidge}, \citenamefont {Burbidge}, \citenamefont {Fowler},\ and\
  \citenamefont {Hoyle}}]{burbidge57}%
  \BibitemOpen
  \bibfield  {author} {\bibinfo {author} {\bibfnamefont {G.}~\bibnamefont
  {Burbidge}}, \bibinfo {author} {\bibfnamefont {R.}~\bibnamefont {Burbidge}},
  \bibinfo {author} {\bibfnamefont {W.}~\bibnamefont {Fowler}}, \ and\ \bibinfo
  {author} {\bibfnamefont {F.}~\bibnamefont {Hoyle}},\ }\href@noop {}
  {\bibfield  {journal} {\bibinfo  {journal} {Rev. Mod. Phys.}\ }\textbf
  {\bibinfo {volume} {29}},\ \bibinfo {pages} {547} (\bibinfo {year}
  {1957})}%
\bibitem [{\citenamefont {Cameron}(1957)}]{cameron57}%
  \BibitemOpen
  \bibfield  {author} {\bibinfo {author} {\bibfnamefont {A.~G.~W.}\
  \bibnamefont {Cameron}},\ }\href@noop {} {\bibfield  {journal} {\bibinfo
  {journal} {Chalk River Rept.}\ }\textbf {\bibinfo {volume} {CRL-41}}
  (\bibinfo {year} {1957})}%
\bibitem [{\citenamefont {Takahashi}\ \emph {et~al.}(1994)\citenamefont
  {Takahashi}, \citenamefont {Witti},\ and\ \citenamefont
  {Janka}}]{takahashi94}%
  \BibitemOpen
  \bibfield  {author} {\bibinfo {author} {\bibfnamefont {K.}~\bibnamefont
  {Takahashi}}, \bibinfo {author} {\bibfnamefont {J.}~\bibnamefont {Witti}}, \
  and\ \bibinfo {author} {\bibfnamefont {H.-T.}\ \bibnamefont {Janka}},\
  }\href@noop {} {\bibfield  {journal} {\bibinfo  {journal} {A\&A}\ }\textbf
  {\bibinfo {volume} {286}},\ \bibinfo {pages} {857} (\bibinfo {year}
  {1994})}%
\bibitem [{\citenamefont {Woosley}\ \emph {et~al.}(1994)\citenamefont
  {Woosley}, \citenamefont {Wilson}, \citenamefont {Mathews}, \citenamefont
  {Hoffman},\ and\ \citenamefont {Meyer}}]{woosley94}%
  \BibitemOpen
  \bibfield  {author} {\bibinfo {author} {\bibfnamefont {S.~E.}\ \bibnamefont
  {Woosley}}, \bibinfo {author} {\bibfnamefont {J.~R.}\ \bibnamefont {Wilson}},
  \bibinfo {author} {\bibfnamefont {G.~J.}\ \bibnamefont {Mathews}}, \bibinfo
  {author} {\bibfnamefont {R.~D.}\ \bibnamefont {Hoffman}}, \ and\ \bibinfo
  {author} {\bibfnamefont {B.~S.}\ \bibnamefont {Meyer}},\ }\href@noop {}
  {\bibfield  {journal} {\bibinfo  {journal} {ApJ}\ }\textbf {\bibinfo {volume}
  {433}},\ \bibinfo {pages} {229} (\bibinfo {year} {1994})}
  %
\bibitem [{\citenamefont {Hoffman}\ \emph {et~al.}(1997)\citenamefont
  {Hoffman}, \citenamefont {Woosley},\ and\ \citenamefont {Qian}}]{hoffman97}%
  \BibitemOpen
  \bibfield  {author} {\bibinfo {author} {\bibfnamefont {R.~D.}\ \bibnamefont
  {Hoffman}}, \bibinfo {author} {\bibfnamefont {S.~E.}\ \bibnamefont
  {Woosley}}, \ and\ \bibinfo {author} {\bibfnamefont {Y.-Z.}\ \bibnamefont
  {Qian}},\ }\href@noop {} {\bibfield  {journal} {\bibinfo  {journal} {ApJ}\
  }\textbf {\bibinfo {volume} {482}},\ \bibinfo {pages} {951} (\bibinfo {year}
  {1997})}%
\bibitem [{\citenamefont {{Freiburghaus}}\ \emph
  {et~al.}(1999{\natexlab{a}})\citenamefont {{Freiburghaus}}, \citenamefont
  {{Rembges}}, \citenamefont {{Rauscher}}, \citenamefont {{Kolbe}},
  \citenamefont {{Thielemann}}, \citenamefont {{Kratz}}, \citenamefont
  {{Pfeiffer}},\ and\ \citenamefont {{Cowan}}}]{freiburghaus99a}%
  \BibitemOpen
  \bibfield  {author} {\bibinfo {author} {\bibfnamefont {C.}~\bibnamefont
  {{Freiburghaus}}}, \bibinfo {author} {\bibfnamefont {J.-F.}\ \bibnamefont
  {{Rembges}}}, \bibinfo {author} {\bibfnamefont {T.}~\bibnamefont
  {{Rauscher}}}, \bibinfo {author} {\bibfnamefont {E.}~\bibnamefont {{Kolbe}}},
  \bibinfo {author} {\bibfnamefont {F.-K.}\ \bibnamefont {{Thielemann}}},
  \bibinfo {author} {\bibfnamefont {K.-L.}\ \bibnamefont {{Kratz}}}, \bibinfo
  {author} {\bibfnamefont {B.}~\bibnamefont {{Pfeiffer}}}, \ and\ \bibinfo
  {author} {\bibfnamefont {J.~J.}\ \bibnamefont {{Cowan}}},\ }\href {\doibase
  10.1086/307072} {\bibfield  {journal} {\bibinfo  {journal} {\apj}\ }\textbf
  {\bibinfo {volume} {516}},\ \bibinfo {pages} {381} (\bibinfo {year}
  {1999}{\natexlab{a}})}%
\bibitem [{\citenamefont {{Farouqi}}\ \emph {et~al.}(2010)\citenamefont
  {{Farouqi}}, \citenamefont {{Kratz}}, \citenamefont {{Pfeiffer}},
  \citenamefont {{Rauscher}}, \citenamefont {{Thielemann}},\ and\ \citenamefont
  {{Truran}}}]{farouqi10}%
  \BibitemOpen
  \bibfield  {author} {\bibinfo {author} {\bibfnamefont {K.}~\bibnamefont
  {{Farouqi}}}, \bibinfo {author} {\bibfnamefont {K.-L.}\ \bibnamefont
  {{Kratz}}}, \bibinfo {author} {\bibfnamefont {B.}~\bibnamefont {{Pfeiffer}}},
  \bibinfo {author} {\bibfnamefont {T.}~\bibnamefont {{Rauscher}}}, \bibinfo
  {author} {\bibfnamefont {F.-K.}\ \bibnamefont {{Thielemann}}}, \ and\
  \bibinfo {author} {\bibfnamefont {J.~W.}\ \bibnamefont {{Truran}}},\ }\href
  {\doibase 10.1088/0004-637X/712/2/1359} {\bibfield  {journal} {\bibinfo
  {journal} {ApJ}\ }\textbf {\bibinfo {volume} {712}},\ \bibinfo {pages} {1359}
  (\bibinfo {year} {2010})},\ \Eprint {http://arxiv.org/abs/1002.2346}
  {arXiv:1002.2346 [astro-ph.SR]} %
\bibitem [{\citenamefont {{Arcones}}\ and\ \citenamefont
  {{Thielemann}}(2013)}]{arcones13a}%
  \BibitemOpen
  \bibfield  {author} {\bibinfo {author} {\bibfnamefont {A.}~\bibnamefont
  {{Arcones}}}\ and\ \bibinfo {author} {\bibfnamefont {F.-K.}\ \bibnamefont
  {{Thielemann}}},\ }\href {\doibase 10.1088/0954-3899/40/1/013201} {\bibfield
  {journal} {\bibinfo  {journal} {Journal of Physics G Nuclear Physics}\
  }\textbf {\bibinfo {volume} {40}},\ \bibinfo {eid} {013201} (\bibinfo {year}
  {2013})},\ \Eprint {http://arxiv.org/abs/1207.2527} {arXiv:1207.2527
  [astro-ph.SR]} %
\bibitem [{\citenamefont {Lattimer}\ and\ \citenamefont
  {Schramm}(1974)}]{lattimer74}%
  \BibitemOpen
  \bibfield  {author} {\bibinfo {author} {\bibfnamefont {J.~M.}\ \bibnamefont
  {Lattimer}}\ and\ \bibinfo {author} {\bibfnamefont {D.~N.}\ \bibnamefont
  {Schramm}},\ }\href@noop {} {\bibfield  {journal} {\bibinfo  {journal} {ApJ,
  (Letters)}\ }\textbf {\bibinfo {volume} {192}},\ \bibinfo {pages} {L145}
  (\bibinfo {year} {1974})}%
\bibitem [{\citenamefont {Rosswog}\ \emph {et~al.}(1999)\citenamefont
  {Rosswog}, \citenamefont {Liebend\"orfer}, \citenamefont {Thielemann},
  \citenamefont {Davies}, \citenamefont {Benz},\ and\ \citenamefont
  {Piran}}]{rosswog99}%
  \BibitemOpen
  \bibfield  {author} {\bibinfo {author} {\bibfnamefont {S.}~\bibnamefont
  {Rosswog}}, \bibinfo {author} {\bibfnamefont {M.}~\bibnamefont
  {Liebend\"orfer}}, \bibinfo {author} {\bibfnamefont {F.-K.}\ \bibnamefont
  {Thielemann}}, \bibinfo {author} {\bibfnamefont {M.}~\bibnamefont {Davies}},
  \bibinfo {author} {\bibfnamefont {W.}~\bibnamefont {Benz}}, \ and\ \bibinfo
  {author} {\bibfnamefont {T.}~\bibnamefont {Piran}},\ }\href@noop {}
  {\bibfield  {journal} {\bibinfo  {journal} {A \&\ A}\ }\textbf {\bibinfo
  {volume} {341}},\ \bibinfo {pages} {499} (\bibinfo {year}
  {1999})}%
\bibitem [{\citenamefont {{Freiburghaus}}\ \emph
  {et~al.}(1999{\natexlab{b}})\citenamefont {{Freiburghaus}}, \citenamefont
  {{Rosswog}},\ and\ \citenamefont {{Thielemann}}}]{freiburghaus99b}%
  \BibitemOpen
  \bibfield  {author} {\bibinfo {author} {\bibfnamefont {C.}~\bibnamefont
  {{Freiburghaus}}}, \bibinfo {author} {\bibfnamefont {S.}~\bibnamefont
  {{Rosswog}}}, \ and\ \bibinfo {author} {\bibfnamefont {F.-K.}\ \bibnamefont
  {{Thielemann}}},\ }\href {\doibase 10.1086/312343} {\bibfield  {journal}
  {\bibinfo  {journal} {\apjl}\ }\textbf {\bibinfo {volume} {525}},\ \bibinfo
  {pages} {L121} (\bibinfo {year} {1999}{\natexlab{b}})}%
\bibitem [{\citenamefont {{Li}}\ and\ \citenamefont
  {{Paczy{\'n}ski}}(1998)}]{LP98}%
  \BibitemOpen
  \bibfield  {author} {\bibinfo {author} {\bibfnamefont {L.}~\bibnamefont
  {{Li}}}\ and\ \bibinfo {author} {\bibfnamefont {B.}~\bibnamefont
  {{Paczy{\'n}ski}}},\ }\href {\doibase 10.1086/311680} {\bibfield  {journal}
  {\bibinfo  {journal} {\apjl}\ }\textbf {\bibinfo {volume} {507}},\ \bibinfo
  {pages} {L59} (\bibinfo {year} {1998})},\ \Eprint
  {http://arxiv.org/abs/arXiv:astro-ph/9807272} {arXiv:astro-ph/9807272}
  %
\bibitem [{\citenamefont {{Rosswog}}\ \emph {et~al.}(2000)\citenamefont
  {{Rosswog}}, \citenamefont {{Davies}}, \citenamefont {{Thielemann}},\ and\
  \citenamefont {{Piran}}}]{rosswog00}%
  \BibitemOpen
  \bibfield  {author} {\bibinfo {author} {\bibfnamefont {S.}~\bibnamefont
  {{Rosswog}}}, \bibinfo {author} {\bibfnamefont {M.~B.}\ \bibnamefont
  {{Davies}}}, \bibinfo {author} {\bibfnamefont {F.-K.}\ \bibnamefont
  {{Thielemann}}}, \ and\ \bibinfo {author} {\bibfnamefont {T.}~\bibnamefont
  {{Piran}}},\ }\href@noop {} {\bibfield  {journal} {\bibinfo  {journal}
  {A\&A}\ }\textbf {\bibinfo {volume} {360}},\ \bibinfo {pages} {171} (\bibinfo
  {year} {2000})}%
\bibitem [{\citenamefont {{Ruffert}}\ and\ \citenamefont
  {{Janka}}(2001)}]{ruffert01}%
  \BibitemOpen
  \bibfield  {author} {\bibinfo {author} {\bibfnamefont {M.}~\bibnamefont
  {{Ruffert}}}\ and\ \bibinfo {author} {\bibfnamefont {H.-T.}\ \bibnamefont
  {{Janka}}},\ }\href {\doibase 10.1051/0004-6361:20011453} {\bibfield
  {journal} {\bibinfo  {journal} {A\&A}\ }\textbf {\bibinfo {volume} {380}},\
  \bibinfo {pages} {544} (\bibinfo {year} {2001})}%
\bibitem [{\citenamefont {{Rosswog}}(2005)}]{rosswog05a}%
  \BibitemOpen
  \bibfield  {author} {\bibinfo {author} {\bibfnamefont {S.}~\bibnamefont
  {{Rosswog}}},\ }\href {\doibase 10.1086/497062} {\bibfield  {journal}
  {\bibinfo  {journal} {ApJ}\ }\textbf {\bibinfo {volume} {634}},\ \bibinfo
  {pages} {1202} (\bibinfo {year} {2005})}%
\bibitem [{\citenamefont {{Oechslin}}\ \emph {et~al.}(2007)\citenamefont
  {{Oechslin}}, \citenamefont {{Janka}},\ and\ \citenamefont
  {{Marek}}}]{oechslin07a}%
  \BibitemOpen
  \bibfield  {author} {\bibinfo {author} {\bibfnamefont {R.}~\bibnamefont
  {{Oechslin}}}, \bibinfo {author} {\bibfnamefont {H.}~\bibnamefont {{Janka}}},
  \ and\ \bibinfo {author} {\bibfnamefont {A.}~\bibnamefont {{Marek}}},\ }\href
  {\doibase 10.1051/0004-6361:20066682} {\bibfield  {journal} {\bibinfo
  {journal} {A \& A}\ }\textbf {\bibinfo {volume} {467}},\ \bibinfo {pages}
  {395} (\bibinfo {year} {2007})},\ \Eprint
  {http://arxiv.org/abs/arXiv:astro-ph/0611047} {arXiv:astro-ph/0611047}
  %
\bibitem [{\citenamefont {{Surman}}\ \emph {et~al.}(2008)\citenamefont
  {{Surman}}, \citenamefont {{McLaughlin}}, \citenamefont {{Ruffert}},
  \citenamefont {{Janka}},\ and\ \citenamefont {{Hix}}}]{surman08}%
  \BibitemOpen
  \bibfield  {author} {\bibinfo {author} {\bibfnamefont {R.}~\bibnamefont
  {{Surman}}}, \bibinfo {author} {\bibfnamefont {G.~C.}\ \bibnamefont
  {{McLaughlin}}}, \bibinfo {author} {\bibfnamefont {M.}~\bibnamefont
  {{Ruffert}}}, \bibinfo {author} {\bibfnamefont {H.}~\bibnamefont {{Janka}}},
  \ and\ \bibinfo {author} {\bibfnamefont {W.~R.}\ \bibnamefont {{Hix}}},\
  }\href {\doibase 10.1086/589507} {\bibfield  {journal} {\bibinfo  {journal}
  {ApJL}\ }\textbf {\bibinfo {volume} {679}},\ \bibinfo {pages} {L117}
  (\bibinfo {year} {2008})},\ \Eprint {http://arxiv.org/abs/0803.1785}
  {arXiv:0803.1785} %
\bibitem [{\citenamefont {{Metzger}}\ \emph {et~al.}(2008)\citenamefont
  {{Metzger}}, \citenamefont {{Piro}},\ and\ \citenamefont
  {{Quataert}}}]{metzger08}%
  \BibitemOpen
  \bibfield  {author} {\bibinfo {author} {\bibfnamefont {B.~D.}\ \bibnamefont
  {{Metzger}}}, \bibinfo {author} {\bibfnamefont {A.~L.}\ \bibnamefont
  {{Piro}}}, \ and\ \bibinfo {author} {\bibfnamefont {E.}~\bibnamefont
  {{Quataert}}},\ }\href {\doibase 10.1111/j.1365-2966.2008.13789.x} {\bibfield
   {journal} {\bibinfo  {journal} {MNRAS}\ }\textbf {\bibinfo {volume} {390}},\
  \bibinfo {pages} {781} (\bibinfo {year} {2008})},\ \Eprint
  {http://arxiv.org/abs/0805.4415} {arXiv:0805.4415} %
\bibitem [{\citenamefont {{Beloborodov}}(2008)}]{beloborodov08}%
  \BibitemOpen
  \bibfield  {author} {\bibinfo {author} {\bibfnamefont {A.~M.}\ \bibnamefont
  {{Beloborodov}}},\ }in\ \href {\doibase 10.1063/1.3002509}
  {American Institute of Physics Conference Series}, Vol.\ \bibinfo
  {volume} {1054},\ \bibinfo {editor} {edited by\ \bibinfo {editor}
  {\bibnamefont {{M.~Axelsson}}}}\ (\bibinfo {year} {2008})\ pp.\ \bibinfo
  {pages} {51--70},\ \Eprint {http://arxiv.org/abs/0810.2690} {arXiv:0810.2690}
  %
\bibitem [{\citenamefont {{Lee}}\ \emph {et~al.}(2009)\citenamefont {{Lee}},
  \citenamefont {{Ramirez-Ruiz}},\ and\ \citenamefont
  {{L{\'o}pez-C{\'a}mara}}}]{lee09}%
  \BibitemOpen
  \bibfield  {author} {\bibinfo {author} {\bibfnamefont {W.~H.}\ \bibnamefont
  {{Lee}}}, \bibinfo {author} {\bibfnamefont {E.}~\bibnamefont
  {{Ramirez-Ruiz}}}, \ and\ \bibinfo {author} {\bibfnamefont {D.}~\bibnamefont
  {{L{\'o}pez-C{\'a}mara}}},\ }\href {\doibase 10.1088/0004-637X/699/2/L93}
  {\bibfield  {journal} {\bibinfo  {journal} {ApJL}\ }\textbf {\bibinfo
  {volume} {699}},\ \bibinfo {pages} {L93} (\bibinfo {year} {2009})},\ \Eprint
  {http://arxiv.org/abs/0904.3752} {arXiv:0904.3752 [astro-ph.HE]} 
  %
\bibitem [{\citenamefont {{Roberts}}\ \emph {et~al.}(2011)\citenamefont
  {{Roberts}}, \citenamefont {{Kasen}}, \citenamefont {{Lee}},\ and\
  \citenamefont {{Ramirez-Ruiz}}}]{roberts11}%
  \BibitemOpen
  \bibfield  {author} {\bibinfo {author} {\bibfnamefont {L.~F.}\ \bibnamefont
  {{Roberts}}}, \bibinfo {author} {\bibfnamefont {D.}~\bibnamefont {{Kasen}}},
  \bibinfo {author} {\bibfnamefont {W.~H.}\ \bibnamefont {{Lee}}}, \ and\
  \bibinfo {author} {\bibfnamefont {E.}~\bibnamefont {{Ramirez-Ruiz}}},\ }\href
  {\doibase 10.1088/2041-8205/736/1/L21} {\bibfield  {journal} {\bibinfo
  {journal} {ApJL}\ }\textbf {\bibinfo {volume} {736}},\ \bibinfo {pages}
  {L21+} (\bibinfo {year} {2011})},\ \Eprint {http://arxiv.org/abs/1104.5504}
  {arXiv:1104.5504 [astro-ph.HE]} %
\bibitem [{\citenamefont {{Goriely}}\ \emph {et~al.}(2011)\citenamefont
  {{Goriely}}, \citenamefont {{Bauswein}},\ and\ \citenamefont
  {{Janka}}}]{goriely11a}%
  \BibitemOpen
  \bibfield  {author} {\bibinfo {author} {\bibfnamefont {S.}~\bibnamefont
  {{Goriely}}}, \bibinfo {author} {\bibfnamefont {A.}~\bibnamefont
  {{Bauswein}}}, \ and\ \bibinfo {author} {\bibfnamefont {H.-T.}\ \bibnamefont
  {{Janka}}},\ }\href {\doibase 10.1088/2041-8205/738/2/L32} {\bibfield
  {journal} {\bibinfo  {journal} {ApJL}\ }\textbf {\bibinfo {volume} {738}},\
  \bibinfo {pages} {L32} (\bibinfo {year} {2011})},\ \Eprint
  {http://arxiv.org/abs/1107.0899} {arXiv:1107.0899 [astro-ph.SR]} 
  %
\bibitem [{\citenamefont {{Korobkin}}\ \emph
  {et~al.}(2012{\natexlab{a}})\citenamefont {{Korobkin}}, \citenamefont
  {{Rosswog}}, \citenamefont {{Arcones}},\ and\ \citenamefont
  {{Winteler}}}]{korobkin12a}%
  \BibitemOpen
  \bibfield  {author} {\bibinfo {author} {\bibfnamefont {O.}~\bibnamefont
  {{Korobkin}}}, \bibinfo {author} {\bibfnamefont {S.}~\bibnamefont
  {{Rosswog}}}, \bibinfo {author} {\bibfnamefont {A.}~\bibnamefont
  {{Arcones}}}, \ and\ \bibinfo {author} {\bibfnamefont {C.}~\bibnamefont
  {{Winteler}}},\ }\href {\doibase 10.1111/j.1365-2966.2012.21859.x} {\bibfield
   {journal} {\bibinfo  {journal} {MNRAS}\ }\textbf {\bibinfo {volume} {426}},\
  \bibinfo {pages} {1940} (\bibinfo {year} {2012}{\natexlab{a}})},\ \Eprint
  {http://arxiv.org/abs/1206.2379} {arXiv:1206.2379 [astro-ph.SR]} 
  %
\bibitem [{\citenamefont {{Metzger}}\ and\ \citenamefont
  {{Berger}}(2012)}]{metzger12a}%
  \BibitemOpen
  \bibfield  {author} {\bibinfo {author} {\bibfnamefont {B.~D.}\ \bibnamefont
  {{Metzger}}}\ and\ \bibinfo {author} {\bibfnamefont {E.}~\bibnamefont
  {{Berger}}},\ }\href {\doibase 10.1088/0004-637X/746/1/48} {\bibfield
  {journal} {\bibinfo  {journal} {ApJ}\ }\textbf {\bibinfo {volume} {746}},\
  \bibinfo {eid} {48} (\bibinfo {year} {2012})},\ \Eprint
  {http://arxiv.org/abs/1108.6056} {arXiv:1108.6056 [astro-ph.HE]} 
  %
\bibitem [{\citenamefont {{Caballero}}\ \emph {et~al.}(2012)\citenamefont
  {{Caballero}}, \citenamefont {{McLaughlin}},\ and\ \citenamefont
  {{Surman}}}]{caballero12}%
  \BibitemOpen
  \bibfield  {author} {\bibinfo {author} {\bibfnamefont {O.~L.}\ \bibnamefont
  {{Caballero}}}, \bibinfo {author} {\bibfnamefont {G.~C.}\ \bibnamefont
  {{McLaughlin}}}, \ and\ \bibinfo {author} {\bibfnamefont {R.}~\bibnamefont
  {{Surman}}},\ }\href {\doibase 10.1088/0004-637X/745/2/170} {\bibfield
  {journal} {\bibinfo  {journal} {ApJ}\ }\textbf {\bibinfo {volume} {745}},\
  \bibinfo {eid} {170} (\bibinfo {year} {2012})},\ \Eprint
  {http://arxiv.org/abs/1105.6371} {arXiv:1105.6371 [astro-ph.HE]} 
  %
\bibitem [{\citenamefont {{Wanajo}}\ and\ \citenamefont
  {{Janka}}(2012)}]{wanajo12}%
  \BibitemOpen
  \bibfield  {author} {\bibinfo {author} {\bibfnamefont {S.}~\bibnamefont
  {{Wanajo}}}\ and\ \bibinfo {author} {\bibfnamefont {H.-T.}\ \bibnamefont
  {{Janka}}},\ }\href {\doibase 10.1088/0004-637X/746/2/180} {\bibfield
  {journal} {\bibinfo  {journal} {\apj}\ }\textbf {\bibinfo {volume} {746}},\
  \bibinfo {eid} {180} (\bibinfo {year} {2012})},\ \Eprint
  {http://arxiv.org/abs/1106.6142} {arXiv:1106.6142 [astro-ph.SR]} 
  %
\bibitem [{\citenamefont {{Malkus}}\ \emph {et~al.}(2012)\citenamefont
  {{Malkus}}, \citenamefont {{Kneller}}, \citenamefont {{McLaughlin}},\ and\
  \citenamefont {{Surman}}}]{malkus12}%
  \BibitemOpen
  \bibfield  {author} {\bibinfo {author} {\bibfnamefont {A.}~\bibnamefont
  {{Malkus}}}, \bibinfo {author} {\bibfnamefont {J.~P.}\ \bibnamefont
  {{Kneller}}}, \bibinfo {author} {\bibfnamefont {G.~C.}\ \bibnamefont
  {{McLaughlin}}}, \ and\ \bibinfo {author} {\bibfnamefont {R.}~\bibnamefont
  {{Surman}}},\ }\href {\doibase 10.1103/PhysRevD.86.085015} {\bibfield
  {journal} {\bibinfo  {journal} {Phys. Rev. D}\ }\textbf {\bibinfo {volume}
  {86}},\ \bibinfo {eid} {085015} (\bibinfo {year} {2012})},\ \Eprint
  {http://arxiv.org/abs/1207.6648} {arXiv:1207.6648 [hep-ph]} 
  %
\bibitem [{\citenamefont {{Kasen}}\ \emph {et~al.}(2013)\citenamefont
  {{Kasen}}, \citenamefont {{Badnell}},\ and\ \citenamefont
  {{Barnes}}}]{kasen13a}%
  \BibitemOpen
  \bibfield  {author} {\bibinfo {author} {\bibfnamefont {D.}~\bibnamefont
  {{Kasen}}}, \bibinfo {author} {\bibfnamefont {N.~R.}\ \bibnamefont
  {{Badnell}}}, \ and\ \bibinfo {author} {\bibfnamefont {J.}~\bibnamefont
  {{Barnes}}},\ }\href@noop {} {\bibfield  {journal} {\bibinfo  {journal}
  {ArXiv e-prints}\ } (\bibinfo {year} {2013})},\ \Eprint
  {http://arxiv.org/abs/1303.5788} {arXiv:1303.5788 [astro-ph.HE]} 
  %
\bibitem [{\citenamefont {{Barnes}}\ and\ \citenamefont
  {{Kasen}}(2013)}]{barnes13a}%
  \BibitemOpen
  \bibfield  {author} {\bibinfo {author} {\bibfnamefont {J.}~\bibnamefont
  {{Barnes}}}\ and\ \bibinfo {author} {\bibfnamefont {D.}~\bibnamefont
  {{Kasen}}},\ }\href {\doibase 10.1088/0004-637X/775/1/18} {\bibfield
  {journal} {\bibinfo  {journal} {\apj}\ }\textbf {\bibinfo {volume} {775}},\
  \bibinfo {eid} {18} (\bibinfo {year} {2013})},\ \Eprint
  {http://arxiv.org/abs/1303.5787} {arXiv:1303.5787 [astro-ph.HE]} 
  %
\bibitem [{\citenamefont {{Rosswog}}\ \emph {et~al.}(2013)\citenamefont
  {{Rosswog}}, \citenamefont {{Piran}},\ and\ \citenamefont
  {{Nakar}}}]{rosswog13a}%
  \BibitemOpen
  \bibfield  {author} {\bibinfo {author} {\bibfnamefont {S.}~\bibnamefont
  {{Rosswog}}}, \bibinfo {author} {\bibfnamefont {T.}~\bibnamefont {{Piran}}},
  \ and\ \bibinfo {author} {\bibfnamefont {E.}~\bibnamefont {{Nakar}}},\ }\href
  {\doibase 10.1093/mnras/sts708} {\bibfield  {journal} {\bibinfo  {journal}
  {MNRAS}\ }\textbf {\bibinfo {volume} {430}},\ \bibinfo {pages} {2585}
  (\bibinfo {year} {2013})},\ \Eprint {http://arxiv.org/abs/1204.6240}
  {arXiv:1204.6240 [astro-ph.HE]} %
\bibitem [{\citenamefont {{Rosswog}}(2013)}]{rosswog13b}%
  \BibitemOpen
  \bibfield  {author} {\bibinfo {author} {\bibfnamefont {S.}~\bibnamefont
  {{Rosswog}}},\ }\href@noop {} {\bibfield  {journal} {\bibinfo  {journal}
  {Philosophical Transactions A, arXiv:1210.6549}\ } (\bibinfo {year}
  {2013})},\ \Eprint {http://arxiv.org/abs/1210.6549} {arXiv:1210.6549
  [astro-ph.HE]} %
\bibitem [{\citenamefont {{Bauswein}}\ \emph {et~al.}(2013)\citenamefont
  {{Bauswein}}, \citenamefont {{Goriely}},\ and\ \citenamefont
  {{Janka}}}]{bauswein13a}%
  \BibitemOpen
  \bibfield  {author} {\bibinfo {author} {\bibfnamefont {A.}~\bibnamefont
  {{Bauswein}}}, \bibinfo {author} {\bibfnamefont {S.}~\bibnamefont
  {{Goriely}}}, \ and\ \bibinfo {author} {\bibfnamefont {H.-T.}\ \bibnamefont
  {{Janka}}},\ }\href@noop {} {\bibfield  {journal} {\bibinfo  {journal} {ArXiv
  e-prints}\ } (\bibinfo {year} {2013})},\ \Eprint
  {http://arxiv.org/abs/1302.6530} {arXiv:1302.6530 [astro-ph.SR]} 
  %
\bibitem [{\citenamefont {{Kyutoku}}\ \emph {et~al.}(2013)\citenamefont
  {{Kyutoku}}, \citenamefont {{Ioka}},\ and\ \citenamefont
  {{Shibata}}}]{kyutoku13}%
  \BibitemOpen
  \bibfield  {author} {\bibinfo {author} {\bibfnamefont {K.}~\bibnamefont
  {{Kyutoku}}}, \bibinfo {author} {\bibfnamefont {K.}~\bibnamefont {{Ioka}}}, \
  and\ \bibinfo {author} {\bibfnamefont {M.}~\bibnamefont {{Shibata}}},\ }\href
  {\doibase 10.1103/PhysRevD.88.041503} {\bibfield  {journal} {\bibinfo
  {journal} {\prd}\ }\textbf {\bibinfo {volume} {88}},\ \bibinfo {eid} {041503}
  (\bibinfo {year} {2013})},\ \Eprint {http://arxiv.org/abs/1305.6309}
  {arXiv:1305.6309 [astro-ph.HE]} %
\bibitem [{\citenamefont {{Surman}}\ \emph {et~al.}(2013)\citenamefont
  {{Surman}}, \citenamefont {{Caballero}}, \citenamefont {{McLaughlin}},
  \citenamefont {{Just}},\ and\ \citenamefont {{Janka}}}]{surman13a}%
  \BibitemOpen
  \bibfield  {author} {\bibinfo {author} {\bibfnamefont {R.}~\bibnamefont
  {{Surman}}}, \bibinfo {author} {\bibfnamefont {O.~L.}\ \bibnamefont
  {{Caballero}}}, \bibinfo {author} {\bibfnamefont {G.~C.}\ \bibnamefont
  {{McLaughlin}}}, \bibinfo {author} {\bibfnamefont {O.}~\bibnamefont
  {{Just}}}, \ and\ \bibinfo {author} {\bibfnamefont {H.-T.}\ \bibnamefont
  {{Janka}}},\ }\href@noop {} {\bibfield  {journal} {\bibinfo  {journal} {ArXiv
  e-prints}\ } (\bibinfo {year} {2013})},\ \Eprint
  {http://arxiv.org/abs/1312.1199} {arXiv:1312.1199 [astro-ph.SR]} 
  %
\bibitem [{\citenamefont {{Fernandez}}\ and\ \citenamefont
  {{Metzger}}(2013)}]{fernandez13}%
  \BibitemOpen
  \bibfield  {author} {\bibinfo {author} {\bibfnamefont {R.}~\bibnamefont
  {{Fernandez}}}\ and\ \bibinfo {author} {\bibfnamefont {B.~D.}\ \bibnamefont
  {{Metzger}}},\ }\href@noop {} {\bibfield  {journal} {\bibinfo  {journal}
  {ArXiv e-prints}\ } (\bibinfo {year} {2013})},\ \Eprint
  {http://arxiv.org/abs/1304.6720} {arXiv:1304.6720 [astro-ph.HE]} 
  %
\bibitem [{\citenamefont {{Arcones}}\ \emph {et~al.}(2007)\citenamefont
  {{Arcones}}, \citenamefont {{Janka}},\ and\ \citenamefont
  {{Scheck}}}]{arcones07}%
  \BibitemOpen
  \bibfield  {author} {\bibinfo {author} {\bibfnamefont {A.}~\bibnamefont
  {{Arcones}}}, \bibinfo {author} {\bibfnamefont {H.-T.}\ \bibnamefont
  {{Janka}}}, \ and\ \bibinfo {author} {\bibfnamefont {L.}~\bibnamefont
  {{Scheck}}},\ }\href {\doibase 10.1051/0004-6361:20066983} {\bibfield
  {journal} {\bibinfo  {journal} {\aap}\ }\textbf {\bibinfo {volume} {467}},\
  \bibinfo {pages} {1227} (\bibinfo {year} {2007})},\ \Eprint
  {http://arxiv.org/abs/astro-ph/0612582} {astro-ph/0612582} 
  %
\bibitem [{\citenamefont {{Roberts}}\ \emph {et~al.}(2010)\citenamefont
  {{Roberts}}, \citenamefont {{Woosley}},\ and\ \citenamefont
  {{Hoffman}}}]{roberts10}%
  \BibitemOpen
  \bibfield  {author} {\bibinfo {author} {\bibfnamefont {L.~F.}\ \bibnamefont
  {{Roberts}}}, \bibinfo {author} {\bibfnamefont {S.~E.}\ \bibnamefont
  {{Woosley}}}, \ and\ \bibinfo {author} {\bibfnamefont {R.~D.}\ \bibnamefont
  {{Hoffman}}},\ }\href@noop {} {\bibfield  {journal} {\bibinfo  {journal}
  {ApJ}\ }\textbf {\bibinfo {volume} {722}},\ \bibinfo {pages} {954} (\bibinfo
  {year} {2010})},\ \Eprint {http://arxiv.org/abs/1004.4916} {arXiv:1004.4916
  [astro-ph.HE]} %
\bibitem [{\citenamefont {{Fischer}}\ \emph {et~al.}(2010)\citenamefont
  {{Fischer}}, \citenamefont {{Whitehouse}}, \citenamefont {{Mezzacappa}},
  \citenamefont {{Thielemann}},\ and\ \citenamefont
  {{Liebend{\"o}rfer}}}]{fischer10}%
  \BibitemOpen
  \bibfield  {author} {\bibinfo {author} {\bibfnamefont {T.}~\bibnamefont
  {{Fischer}}}, \bibinfo {author} {\bibfnamefont {S.~C.}\ \bibnamefont
  {{Whitehouse}}}, \bibinfo {author} {\bibfnamefont {A.}~\bibnamefont
  {{Mezzacappa}}}, \bibinfo {author} {\bibfnamefont {F.-K.}\ \bibnamefont
  {{Thielemann}}}, \ and\ \bibinfo {author} {\bibfnamefont {M.}~\bibnamefont
  {{Liebend{\"o}rfer}}},\ }\href {\doibase 10.1051/0004-6361/200913106}
  {\bibfield  {journal} {\bibinfo  {journal} {A \& A}\ }\textbf {\bibinfo
  {volume} {517}},\ \bibinfo {pages} {A80} (\bibinfo {year} {2010})},\ \Eprint
  {http://arxiv.org/abs/0908.1871} {arXiv:0908.1871 [astro-ph.HE]} 
  %
\bibitem [{\citenamefont {{H{\"u}depohl}}\ \emph {et~al.}(2010)\citenamefont
  {{H{\"u}depohl}}, \citenamefont {{M{\"u}ller}}, \citenamefont {{Janka}},
  \citenamefont {{Marek}},\ and\ \citenamefont {{Raffelt}}}]{huedepohl10}%
  \BibitemOpen
  \bibfield  {author} {\bibinfo {author} {\bibfnamefont {L.}~\bibnamefont
  {{H{\"u}depohl}}}, \bibinfo {author} {\bibfnamefont {B.}~\bibnamefont
  {{M{\"u}ller}}}, \bibinfo {author} {\bibfnamefont {H.-T.}\ \bibnamefont
  {{Janka}}}, \bibinfo {author} {\bibfnamefont {A.}~\bibnamefont {{Marek}}}, \
  and\ \bibinfo {author} {\bibfnamefont {G.~G.}\ \bibnamefont {{Raffelt}}},\
  }\href {\doibase 10.1103/PhysRevLett.104.251101} {\bibfield  {journal}
  {\bibinfo  {journal} {Physical Review Letters}\ }\textbf {\bibinfo {volume}
  {104}},\ \bibinfo {eid} {251101} (\bibinfo {year} {2010})},\ \Eprint
  {http://arxiv.org/abs/0912.0260} {arXiv:0912.0260 [astro-ph.SR]} 
  %
\bibitem [{\citenamefont {{Roberts}}\ \emph {et~al.}(2012)\citenamefont
  {{Roberts}}, \citenamefont {{Reddy}},\ and\ \citenamefont
  {{Shen}}}]{roberts12a}%
  \BibitemOpen
  \bibfield  {author} {\bibinfo {author} {\bibfnamefont {L.~F.}\ \bibnamefont
  {{Roberts}}}, \bibinfo {author} {\bibfnamefont {S.}~\bibnamefont {{Reddy}}},
  \ and\ \bibinfo {author} {\bibfnamefont {G.}~\bibnamefont {{Shen}}},\ }\href
  {\doibase 10.1103/PhysRevC.86.065803} {\bibfield  {journal} {\bibinfo
  {journal} {Phys. Rev. C}\ }\textbf {\bibinfo {volume} {86}},\ \bibinfo {eid}
  {065803} (\bibinfo {year} {2012})},\ \Eprint {http://arxiv.org/abs/1205.4066}
  {arXiv:1205.4066 [astro-ph.HE]} %
\bibitem [{\citenamefont {{Martinez-Pinedo}}\ \emph {et~al.}(2012)\citenamefont
  {{Martinez-Pinedo}}, \citenamefont {{Fischer}}, \citenamefont {{Lohs}},\ and\
  \citenamefont {{Huther}}}]{martinez_pinedo12a}%
  \BibitemOpen
  \bibfield  {author} {\bibinfo {author} {\bibfnamefont {G.}~\bibnamefont
  {{Martinez-Pinedo}}}, \bibinfo {author} {\bibfnamefont {T.}~\bibnamefont
  {{Fischer}}}, \bibinfo {author} {\bibfnamefont {A.}~\bibnamefont {{Lohs}}}, \
  and\ \bibinfo {author} {\bibfnamefont {L.}~\bibnamefont {{Huther}}},\ }\href
  {\doibase 10.1103/PhysRevLett.109.251104} {\bibfield  {journal} {\bibinfo
  {journal} {Physical Review Letters}\ }\textbf {\bibinfo {volume} {109}},\
  \bibinfo {eid} {251104} (\bibinfo {year} {2012})},\ \Eprint
  {http://arxiv.org/abs/1205.2793} {arXiv:1205.2793 [astro-ph.HE]} 
  %
\bibitem [{\citenamefont {{Wanajo}}(2013)}]{wanajo13a}%
  \BibitemOpen
  \bibfield  {author} {\bibinfo {author} {\bibfnamefont {S.}~\bibnamefont
  {{Wanajo}}},\ }\href {\doibase 10.1088/2041-8205/770/2/L22} {\bibfield
  {journal} {\bibinfo  {journal} {ApJL}\ }\textbf {\bibinfo {volume} {770}},\
  \bibinfo {eid} {L22} (\bibinfo {year} {2013})},\ \Eprint
  {http://arxiv.org/abs/1305.0371} {arXiv:1305.0371 [astro-ph.SR]} 
  %
\bibitem [{\citenamefont {{Piran}}\ and\ \citenamefont
  {{Shaviv}}(2005)}]{piran05a}%
  \BibitemOpen
  \bibfield  {author} {\bibinfo {author} {\bibfnamefont {T.}~\bibnamefont
  {{Piran}}}\ and\ \bibinfo {author} {\bibfnamefont {N.~J.}\ \bibnamefont
  {{Shaviv}}},\ }\href {\doibase 10.1103/PhysRevLett.94.051102} {\bibfield
  {journal} {\bibinfo  {journal} {Physical Review Letters}\ }\textbf {\bibinfo
  {volume} {94}},\ \bibinfo {eid} {051102} (\bibinfo {year} {2005})},\ \Eprint
  {http://arxiv.org/abs/astro-ph/0409651} {astro-ph/0409651} 
 \bibitem [{\citenamefont {{Winteler}}\ \emph {et~al.}(2012)\citenamefont
  {{Winteler}}, \citenamefont {{K{\"a}ppeli}}, \citenamefont {{Perego}},
  \citenamefont {{Arcones}}, \citenamefont {{Vasset}}, \citenamefont
  {{Nishimura}}, \citenamefont {{Liebend{\"o}rfer}},\ and\ \citenamefont
  {{Thielemann}}}]{winteler12a}%
  \BibitemOpen
  \bibfield  {author} {\bibinfo {author} {\bibfnamefont {C.}~\bibnamefont
  {{Winteler}}}, \bibinfo {author} {\bibfnamefont {R.}~\bibnamefont
  {{K{\"a}ppeli}}}, \bibinfo {author} {\bibfnamefont {A.}~\bibnamefont
  {{Perego}}}, \bibinfo {author} {\bibfnamefont {A.}~\bibnamefont {{Arcones}}},
  \bibinfo {author} {\bibfnamefont {N.}~\bibnamefont {{Vasset}}}, \bibinfo
  {author} {\bibfnamefont {N.}~\bibnamefont {{Nishimura}}}, \bibinfo {author}
  {\bibfnamefont {M.}~\bibnamefont {{Liebend{\"o}rfer}}}, \ and\ \bibinfo
  {author} {\bibfnamefont {F.-K.}\ \bibnamefont {{Thielemann}}},\ }\href
  {\doibase 10.1088/2041-8205/750/1/L22} {\bibfield  {journal} {\bibinfo
  {journal} {\apjl}\ }\textbf {\bibinfo {volume} {750}},\ \bibinfo {eid} {L22}
  (\bibinfo {year} {2012})},\ \Eprint {http://arxiv.org/abs/1203.0616}
  {arXiv:1203.0616 [astro-ph.SR]} %
\bibitem [{\citenamefont {{Qian}}(2000)}]{qian00}%
  \BibitemOpen
  \bibfield  {author} {\bibinfo {author} {\bibfnamefont {Y.-Z.}\ \bibnamefont
  {{Qian}}},\ }\href {\doibase 10.1086/312659} {\bibfield  {journal} {\bibinfo
  {journal} {ApJL}\ }\textbf {\bibinfo {volume} {534}},\ \bibinfo {pages} {L67}
  (\bibinfo {year} {2000})},\ \Eprint
  {http://arxiv.org/abs/arXiv:astro-ph/0003242} {arXiv:astro-ph/0003242}
  %
\bibitem [{\citenamefont {{Argast}}\ \emph {et~al.}(2004)\citenamefont
  {{Argast}}, \citenamefont {{Samland}}, \citenamefont {{Thielemann}},\ and\
  \citenamefont {{Qian}}}]{argast04}%
  \BibitemOpen
  \bibfield  {author} {\bibinfo {author} {\bibfnamefont {D.}~\bibnamefont
  {{Argast}}}, \bibinfo {author} {\bibfnamefont {M.}~\bibnamefont {{Samland}}},
  \bibinfo {author} {\bibfnamefont {F.-K.}\ \bibnamefont {{Thielemann}}}, \
  and\ \bibinfo {author} {\bibfnamefont {Y.-Z.}\ \bibnamefont {{Qian}}},\
  }\href {\doibase 10.1051/0004-6361:20034265} {\bibfield  {journal} {\bibinfo
  {journal} {A\&A}\ }\textbf {\bibinfo {volume} {416}},\ \bibinfo {pages} {997}
  (\bibinfo {year} {2004})}%
\bibitem [{\citenamefont {{Matteucci}}\ \emph {et~al.}(2013)\citenamefont
  {{Matteucci}}, \citenamefont {{Romano}}, \citenamefont {{Arcones}},
  \citenamefont {{Korobkin}},\ and\ \citenamefont {{Rosswog}}}]{matteucci13}%
  \BibitemOpen
  \bibfield  {author} {\bibinfo {author} {\bibfnamefont {F.}~\bibnamefont
  {{Matteucci}}}, \bibinfo {author} {\bibfnamefont {D.}~\bibnamefont
  {{Romano}}}, \bibinfo {author} {\bibfnamefont {A.}~\bibnamefont {{Arcones}}},
  \bibinfo {author} {\bibfnamefont {O.}~\bibnamefont {{Korobkin}}}, \ and\
  \bibinfo {author} {\bibfnamefont {S.}~\bibnamefont {{Rosswog}}},\ }\href@noop
  {} {\bibfield  {journal} {\bibinfo  {journal} {ArXiv e-prints}\ } (\bibinfo
  {year} {2013})},\ \Eprint {http://arxiv.org/abs/1311.6980} {arXiv:1311.6980
  [astro-ph.GA]} %
\bibitem [{\citenamefont {{Kulkarni}}(2005)}]{kulkarni05}%
  \BibitemOpen
  \bibfield  {author} {\bibinfo {author} {\bibfnamefont {S.~R.}\ \bibnamefont
  {{Kulkarni}}},\ }\href@noop {} {\bibfield  {journal} {\bibinfo  {journal}
  {ArXiv Astrophysics e-prints}\ } (\bibinfo {year} {2005})},\ \Eprint
  {http://arxiv.org/abs/arXiv:astro-ph/0510256} {arXiv:astro-ph/0510256}
  %
\bibitem [{\citenamefont {{Metzger}}\ \emph {et~al.}(2010)\citenamefont
  {{Metzger}}, \citenamefont {{Arcones}}, \citenamefont {{Quataert}},\ and\
  \citenamefont {{Martinez-Pinedo}}}]{metzger10a}%
  \BibitemOpen
  \bibfield  {author} {\bibinfo {author} {\bibfnamefont {B.~D.}\ \bibnamefont
  {{Metzger}}}, \bibinfo {author} {\bibfnamefont {A.}~\bibnamefont
  {{Arcones}}}, \bibinfo {author} {\bibfnamefont {E.}~\bibnamefont
  {{Quataert}}}, \ and\ \bibinfo {author} {\bibfnamefont {G.}~\bibnamefont
  {{Martinez-Pinedo}}},\ }\href {\doibase 10.1111/j.1365-2966.2009.16107.x}
  {\bibfield  {journal} {\bibinfo  {journal} {MNRAS}\ }\textbf {\bibinfo
  {volume} {402}},\ \bibinfo {pages} {2771} (\bibinfo {year} {2010})},\ \Eprint
  {http://arxiv.org/abs/0908.0530} {arXiv:0908.0530} %
\bibitem [{\citenamefont {{Piran}}\ \emph {et~al.}(2013)\citenamefont
  {{Piran}}, \citenamefont {{Nakar}},\ and\ \citenamefont
  {{Rosswog}}}]{piran13a}%
  \BibitemOpen
  \bibfield  {author} {\bibinfo {author} {\bibfnamefont {T.}~\bibnamefont
  {{Piran}}}, \bibinfo {author} {\bibfnamefont {E.}~\bibnamefont {{Nakar}}}, \
  and\ \bibinfo {author} {\bibfnamefont {S.}~\bibnamefont {{Rosswog}}},\ }\href
  {\doibase 10.1093/mnras/stt037} {\bibfield  {journal} {\bibinfo  {journal}
  {\mnras}\ }\textbf {\bibinfo {volume} {430}},\ \bibinfo {pages} {2121}
  (\bibinfo {year} {2013})},\ \Eprint {http://arxiv.org/abs/1204.6242}
  {arXiv:1204.6242 [astro-ph.HE]} %
\bibitem [{\citenamefont {{Tanaka}}\ and\ \citenamefont
  {{Hotokezaka}}(2013)}]{tanaka13a}%
  \BibitemOpen
  \bibfield  {author} {\bibinfo {author} {\bibfnamefont {M.}~\bibnamefont
  {{Tanaka}}}\ and\ \bibinfo {author} {\bibfnamefont {K.}~\bibnamefont
  {{Hotokezaka}}},\ }\href@noop {} {\bibfield  {journal} {\bibinfo  {journal}
  {ArXiv e-prints}\ } (\bibinfo {year} {2013})},\ \Eprint
  {http://arxiv.org/abs/1306.3742} {arXiv:1306.3742 [astro-ph.HE]} 
  %
\bibitem [{\citenamefont {{Grossman}}\ \emph {et~al.}(2013)\citenamefont
  {{Grossman}}, \citenamefont {{Korobkin}}, \citenamefont {{Rosswog}},\ and\
  \citenamefont {{Piran}}}]{grossman13a}%
  \BibitemOpen
  \bibfield  {author} {\bibinfo {author} {\bibfnamefont {D.}~\bibnamefont
  {{Grossman}}}, \bibinfo {author} {\bibfnamefont {O.}~\bibnamefont
  {{Korobkin}}}, \bibinfo {author} {\bibfnamefont {S.}~\bibnamefont
  {{Rosswog}}}, \ and\ \bibinfo {author} {\bibfnamefont {T.}~\bibnamefont
  {{Piran}}},\ }\href@noop {} {\bibfield  {journal} {\bibinfo  {journal} {ArXiv
  e-prints}\ } (\bibinfo {year} {2013})},\ \Eprint
  {http://arxiv.org/abs/1307.2943} {arXiv:1307.2943 [astro-ph.HE]} 
  %
\bibitem [{\citenamefont {{Melandri}}\ \emph {et~al.}(2013)\citenamefont
  {{Melandri}}, \citenamefont {{Baumgartner}}, \citenamefont {{Burrows}},
  \citenamefont {{Cummings}}, \citenamefont {{Gehrels}}, \citenamefont
  {{Gronwall}}, \citenamefont {{Page}}, \citenamefont {{Palmer}}, \citenamefont
  {{Starling}},\ and\ \citenamefont {{Ukwatta}}}]{Melandri13}%
  \BibitemOpen
  \bibfield  {author} {\bibinfo {author} {\bibfnamefont {A.}~\bibnamefont
  {{Melandri}}}, \bibinfo {author} {\bibfnamefont {W.~H.}\ \bibnamefont
  {{Baumgartner}}}, \bibinfo {author} {\bibfnamefont {D.~N.}\ \bibnamefont
  {{Burrows}}}, \bibinfo {author} {\bibfnamefont {J.~R.}\ \bibnamefont
  {{Cummings}}}, \bibinfo {author} {\bibfnamefont {N.}~\bibnamefont
  {{Gehrels}}}, \bibinfo {author} {\bibfnamefont {C.}~\bibnamefont
  {{Gronwall}}}, \bibinfo {author} {\bibfnamefont {K.~L.}\ \bibnamefont
  {{Page}}}, \bibinfo {author} {\bibfnamefont {D.~M.}\ \bibnamefont
  {{Palmer}}}, \bibinfo {author} {\bibfnamefont {R.~L.~C.}\ \bibnamefont
  {{Starling}}}, \ and\ \bibinfo {author} {\bibfnamefont {T.~N.}\ \bibnamefont
  {{Ukwatta}}},\ }\href@noop {} {\bibfield  {journal} {\bibinfo  {journal} {GRB
  Coordinates Network}\ }\textbf {\bibinfo {volume} {14735}},\ \bibinfo {pages}
  {1} (\bibinfo {year} {2013})}%
\bibitem [{\citenamefont {{Bromberg}}\ \emph {et~al.}(2013)\citenamefont
  {{Bromberg}}, \citenamefont {{Nakar}}, \citenamefont {{Piran}},\ and\
  \citenamefont {{Sari}}}]{Bromberg13}%
  \BibitemOpen
  \bibfield  {author} {\bibinfo {author} {\bibfnamefont {O.}~\bibnamefont
  {{Bromberg}}}, \bibinfo {author} {\bibfnamefont {E.}~\bibnamefont {{Nakar}}},
  \bibinfo {author} {\bibfnamefont {T.}~\bibnamefont {{Piran}}}, \ and\
  \bibinfo {author} {\bibfnamefont {R.}~\bibnamefont {{Sari}}},\ }\href
  {\doibase 10.1088/0004-637X/764/2/179} {\bibfield  {journal} {\bibinfo
  {journal} {\apj}\ }\textbf {\bibinfo {volume} {764}},\ \bibinfo {eid} {179}
  (\bibinfo {year} {2013})},\ \Eprint {http://arxiv.org/abs/1210.0068}
  {arXiv:1210.0068 [astro-ph.HE]} %
\bibitem [{\citenamefont {{Fong}}\ \emph
  {et~al.}(2013{\natexlab{a}})\citenamefont {{Fong}}, \citenamefont {{Berger}},
  \citenamefont {{Metzger}}, \citenamefont {{Margutti}}, \citenamefont
  {{Chornock}}, \citenamefont {{Migliori}}, \citenamefont {{Foley}},\ and\
  \citenamefont {{Zauderer}}}]{fong13b}%
  \BibitemOpen
  \bibfield  {author} {\bibinfo {author} {\bibfnamefont {W.-f.}\ \bibnamefont
  {{Fong}}}, \bibinfo {author} {\bibfnamefont {E.}~\bibnamefont {{Berger}}},
  \bibinfo {author} {\bibfnamefont {B.~D.}\ \bibnamefont {{Metzger}}}, \bibinfo
  {author} {\bibfnamefont {R.}~\bibnamefont {{Margutti}}}, \bibinfo {author}
  {\bibfnamefont {R.}~\bibnamefont {{Chornock}}}, \bibinfo {author}
  {\bibfnamefont {G.}~\bibnamefont {{Migliori}}}, \bibinfo {author}
  {\bibfnamefont {R.~J.}\ \bibnamefont {{Foley}}}, \ and\ \bibinfo {author}
  {\bibfnamefont {B.~A.}\ \bibnamefont {{Zauderer}}},\ }\href@noop {}
  {\bibfield  {journal} {\bibinfo  {journal} {ArXiv e-prints}\ } (\bibinfo
  {year} {2013}{\natexlab{a}})},\ \Eprint {http://arxiv.org/abs/1309.7479}
  {arXiv:1309.7479 [astro-ph.HE]} %
\bibitem [{\citenamefont {Rosswog}(2007)}]{rosswog07a}%
  \BibitemOpen
  \bibfield  {author} {\bibinfo {author} {\bibfnamefont {S.}~\bibnamefont
  {Rosswog}},\ }\href@noop {} {\bibfield  {journal} {\bibinfo  {journal}
  {MNRAS}\ }\textbf {\bibinfo {volume} {376}},\ \bibinfo {pages} {L48}
  (\bibinfo {year} {2007})}%
\bibitem [{\citenamefont {{Tanvir}}\ \emph {et~al.}(2013)\citenamefont
  {{Tanvir}}, \citenamefont {{Levan}}, \citenamefont {{Fruchter}},
  \citenamefont {{Hjorth}}, \citenamefont {{Hounsell}}, \citenamefont
  {{Wiersema}},\ and\ \citenamefont {{Tunnicliffe}}}]{Tanvir+13}%
  \BibitemOpen
  \bibfield  {author} {\bibinfo {author} {\bibfnamefont {N.~R.}\ \bibnamefont
  {{Tanvir}}}, \bibinfo {author} {\bibfnamefont {A.~J.}\ \bibnamefont
  {{Levan}}}, \bibinfo {author} {\bibfnamefont {A.~S.}\ \bibnamefont
  {{Fruchter}}}, \bibinfo {author} {\bibfnamefont {J.}~\bibnamefont
  {{Hjorth}}}, \bibinfo {author} {\bibfnamefont {R.~A.}\ \bibnamefont
  {{Hounsell}}}, \bibinfo {author} {\bibfnamefont {K.}~\bibnamefont
  {{Wiersema}}}, \ and\ \bibinfo {author} {\bibfnamefont {R.~L.}\ \bibnamefont
  {{Tunnicliffe}}},\ }\href {\doibase 10.1038/nature12505} {\bibfield
  {journal} {\bibinfo  {journal} {\nat}\ }\textbf {\bibinfo {volume} {500}},\
  \bibinfo {pages} {547} (\bibinfo {year} {2013})},\ \Eprint
  {http://arxiv.org/abs/1306.4971} {arXiv:1306.4971 [astro-ph.HE]} 
  %
\bibitem [{\citenamefont {{Berger}}\ \emph {et~al.}(2013)\citenamefont
  {{Berger}}, \citenamefont {{Fong}},\ and\ \citenamefont
  {{Chornock}}}]{Berger+13}%
  \BibitemOpen
  \bibfield  {author} {\bibinfo {author} {\bibfnamefont {E.}~\bibnamefont
  {{Berger}}}, \bibinfo {author} {\bibfnamefont {W.}~\bibnamefont {{Fong}}}, \
  and\ \bibinfo {author} {\bibfnamefont {R.}~\bibnamefont {{Chornock}}},\
  }\href {\doibase 10.1088/2041-8205/774/2/L23} {\bibfield  {journal} {\bibinfo
   {journal} {\apjl}\ }\textbf {\bibinfo {volume} {774}},\ \bibinfo {eid} {L23}
  (\bibinfo {year} {2013})},\ \Eprint {http://arxiv.org/abs/1306.3960}
  {arXiv:1306.3960 [astro-ph.HE]} %
\bibitem [{\citenamefont {{Nakar}}(2007)}]{Nakar07}%
  \BibitemOpen
  \bibfield  {author} {\bibinfo {author} {\bibfnamefont {E.}~\bibnamefont
  {{Nakar}}},\ }\href {\doibase 10.1016/j.physrep.2007.02.005} {\bibfield
  {journal} {\bibinfo  {journal} {\physrep}\ }\textbf {\bibinfo {volume}
  {442}},\ \bibinfo {pages} {166} (\bibinfo {year} {2007})},\ \Eprint
  {http://arxiv.org/abs/arXiv:astro-ph/0701748} {arXiv:astro-ph/0701748}
  %
\bibitem [{\citenamefont {{Fong}}\ \emph
  {et~al.}(2013{\natexlab{b}})\citenamefont {{Fong}}, \citenamefont {{Berger}},
  \citenamefont {{Chornock}}, \citenamefont {{Margutti}}, \citenamefont
  {{Levan}}, \citenamefont {{Tanvir}},\ and\ \citenamefont
  {{Tunnicliffe}}}]{Fong+13}%
  \BibitemOpen
  \bibfield  {author} {\bibinfo {author} {\bibfnamefont {W.}~\bibnamefont
  {{Fong}}}, \bibinfo {author} {\bibfnamefont {E.}~\bibnamefont {{Berger}}},
  \bibinfo {author} {\bibfnamefont {R.}~\bibnamefont {{Chornock}}}, \bibinfo
  {author} {\bibfnamefont {R.}~\bibnamefont {{Margutti}}}, \bibinfo {author}
  {\bibfnamefont {A.~J.}\ \bibnamefont {{Levan}}}, \bibinfo {author}
  {\bibfnamefont {N.~R.}\ \bibnamefont {{Tanvir}}}, \ and\ \bibinfo {author}
  {\bibfnamefont {R.~L.}\ \bibnamefont {{Tunnicliffe}}},\ }\href {\doibase
  10.1088/0004-637X/769/1/56} {\bibfield  {journal} {\bibinfo  {journal}
  {\apj}\ }\textbf {\bibinfo {volume} {769}},\ \bibinfo {eid} {56} (\bibinfo
  {year} {2013}{\natexlab{b}})},\ \Eprint {http://arxiv.org/abs/1302.3221}
  {arXiv:1302.3221 [astro-ph.HE]} %
\bibitem [{\citenamefont {{Guetta}}\ and\ \citenamefont
  {{Piran}}(2006)}]{guetta06}%
  \BibitemOpen
  \bibfield  {author} {\bibinfo {author} {\bibfnamefont {D.}~\bibnamefont
  {{Guetta}}}\ and\ \bibinfo {author} {\bibfnamefont {T.}~\bibnamefont
  {{Piran}}},\ }\href {\doibase 10.1051/0004-6361:20054498} {\bibfield
  {journal} {\bibinfo  {journal} {\aap}\ }\textbf {\bibinfo {volume} {453}},\
  \bibinfo {pages} {823} (\bibinfo {year} {2006})},\ \Eprint
  {http://arxiv.org/abs/astro-ph/0511239} {astro-ph/0511239} 
  %
\bibitem [{\citenamefont {{Nakar}}\ \emph {et~al.}(2006)\citenamefont
  {{Nakar}}, \citenamefont {{Gal-Yam}},\ and\ \citenamefont {{Fox}}}]{NGF06}%
  \BibitemOpen
  \bibfield  {author} {\bibinfo {author} {\bibfnamefont {E.}~\bibnamefont
  {{Nakar}}}, \bibinfo {author} {\bibfnamefont {A.}~\bibnamefont {{Gal-Yam}}},
  \ and\ \bibinfo {author} {\bibfnamefont {D.~B.}\ \bibnamefont {{Fox}}},\
  }\href {\doibase 10.1086/505855} {\bibfield  {journal} {\bibinfo  {journal}
  {\apj}\ }\textbf {\bibinfo {volume} {650}},\ \bibinfo {pages} {281} (\bibinfo
  {year} {2006})},\ \Eprint {http://arxiv.org/abs/arXiv:astro-ph/0511254}
  {arXiv:astro-ph/0511254} %
\bibitem [{\citenamefont {{Korobkin}}\ \emph
  {et~al.}(2012{\natexlab{b}})\citenamefont {{Korobkin}}, \citenamefont
  {{Rosswog}}, \citenamefont {{Arcones}},\ and\ \citenamefont
  {{Winteler}}}]{korobkin12}%
  \BibitemOpen
  \bibfield  {author} {\bibinfo {author} {\bibfnamefont {O.}~\bibnamefont
  {{Korobkin}}}, \bibinfo {author} {\bibfnamefont {S.}~\bibnamefont
  {{Rosswog}}}, \bibinfo {author} {\bibfnamefont {A.}~\bibnamefont
  {{Arcones}}}, \ and\ \bibinfo {author} {\bibfnamefont {C.}~\bibnamefont
  {{Winteler}}},\ }\href {\doibase 10.1111/j.1365-2966.2012.21859.x} {\bibfield
   {journal} {\bibinfo  {journal} {\mnras}\ }\textbf {\bibinfo {volume}
  {426}},\ \bibinfo {pages} {1940} (\bibinfo {year} {2012}{\natexlab{b}})},\
  \Eprint {http://arxiv.org/abs/1206.2379} {arXiv:1206.2379 [astro-ph.SR]}
  %
\bibitem [{\citenamefont {{Kappeler}}\ \emph {et~al.}(1989)\citenamefont
  {{Kappeler}}, \citenamefont {{Beer}},\ and\ \citenamefont
  {{Wisshak}}}]{kappler89}%
  \BibitemOpen
  \bibfield  {author} {\bibinfo {author} {\bibfnamefont {F.}~\bibnamefont
  {{Kappeler}}}, \bibinfo {author} {\bibfnamefont {H.}~\bibnamefont {{Beer}}},
  \ and\ \bibinfo {author} {\bibfnamefont {K.}~\bibnamefont {{Wisshak}}},\
  }\href {\doibase 10.1088/0034-4885/52/8/002} {\bibfield  {journal} {\bibinfo
  {journal} {Reports on Progress in Physics}\ }\textbf {\bibinfo {volume}
  {52}},\ \bibinfo {pages} {945} (\bibinfo {year} {1989})}
  %
\bibitem [{\citenamefont {{Wanderman}}\ and\ \citenamefont
  {{Piran}}(2014)}]{Wanderman13}%
  \BibitemOpen
  \bibfield  {author} {\bibinfo {author} {\bibfnamefont {D.}~\bibnamefont
  {{Wanderman}}}\ and\ \bibinfo {author} {\bibfnamefont {T.}~\bibnamefont
  {{Piran}}},\ }\href@noop {} {\bibfield  {journal} {\bibinfo  {journal} {in
  preparation}\ }\textbf {\bibinfo {volume} {498}} (\bibinfo {year}
  {2014})}%
\bibitem [{\citenamefont {{Guetta}}\ and\ \citenamefont
  {{Stella}}(2009)}]{GS09}%
  \BibitemOpen
  \bibfield  {author} {\bibinfo {author} {\bibfnamefont {D.}~\bibnamefont
  {{Guetta}}}\ and\ \bibinfo {author} {\bibfnamefont {L.}~\bibnamefont
  {{Stella}}},\ }\href {\doibase 10.1051/0004-6361:200810493} {\bibfield
  {journal} {\bibinfo  {journal} {\aap}\ }\textbf {\bibinfo {volume} {498}},\
  \bibinfo {pages} {329} (\bibinfo {year} {2009})},\ \Eprint
  {http://arxiv.org/abs/0811.0684} {arXiv:0811.0684} %
\bibitem [{\citenamefont {{Narayan}}\ \emph {et~al.}(1991)\citenamefont
  {{Narayan}}, \citenamefont {{Piran}},\ and\ \citenamefont {{Shemi}}}]{NPS91}%
  \BibitemOpen
  \bibfield  {author} {\bibinfo {author} {\bibfnamefont {R.}~\bibnamefont
  {{Narayan}}}, \bibinfo {author} {\bibfnamefont {T.}~\bibnamefont {{Piran}}},
  \ and\ \bibinfo {author} {\bibfnamefont {A.}~\bibnamefont {{Shemi}}},\ }\href
  {\doibase 10.1086/186143} {\bibfield  {journal} {\bibinfo  {journal} {\apjl}\
  }\textbf {\bibinfo {volume} {379}},\ \bibinfo {pages} {L17} (\bibinfo {year}
  {1991})}%
\bibitem [{\citenamefont {{Phinney}}(1991)}]{Phinney91}%
  \BibitemOpen
  \bibfield  {author} {\bibinfo {author} {\bibfnamefont {E.~S.}\ \bibnamefont
  {{Phinney}}},\ }\href {\doibase 10.1086/186163} {\bibfield  {journal}
  {\bibinfo  {journal} {\apjl}\ }\textbf {\bibinfo {volume} {380}},\ \bibinfo
  {pages} {L17} (\bibinfo {year} {1991})}%
\bibitem [{\citenamefont {{Kalogera}}\ \emph
  {et~al.}(2004{\natexlab{a}})\citenamefont {{Kalogera}}, \citenamefont
  {{Kim}}, \citenamefont {{Lorimer}}, \citenamefont {{Burgay}}, \citenamefont
  {{D'Amico}}, \citenamefont {{Possenti}}, \citenamefont {{Manchester}},
  \citenamefont {{Lyne}}, \citenamefont {{Joshi}}, \citenamefont
  {{McLaughlin}}, \citenamefont {{Kramer}}, \citenamefont {{Sarkissian}},\ and\
  \citenamefont {{Camilo}}}]{KalEtal04}%
  \BibitemOpen
  \bibfield  {author} {\bibinfo {author} {\bibfnamefont {V.}~\bibnamefont
  {{Kalogera}}}, \bibinfo {author} {\bibfnamefont {C.}~\bibnamefont {{Kim}}},
  \bibinfo {author} {\bibfnamefont {D.~R.}\ \bibnamefont {{Lorimer}}}, \bibinfo
  {author} {\bibfnamefont {M.}~\bibnamefont {{Burgay}}}, \bibinfo {author}
  {\bibfnamefont {N.}~\bibnamefont {{D'Amico}}}, \bibinfo {author}
  {\bibfnamefont {A.}~\bibnamefont {{Possenti}}}, \bibinfo {author}
  {\bibfnamefont {R.~N.}\ \bibnamefont {{Manchester}}}, \bibinfo {author}
  {\bibfnamefont {A.~G.}\ \bibnamefont {{Lyne}}}, \bibinfo {author}
  {\bibfnamefont {B.~C.}\ \bibnamefont {{Joshi}}}, \bibinfo {author}
  {\bibfnamefont {M.~A.}\ \bibnamefont {{McLaughlin}}}, \bibinfo {author}
  {\bibfnamefont {M.}~\bibnamefont {{Kramer}}}, \bibinfo {author}
  {\bibfnamefont {J.~M.}\ \bibnamefont {{Sarkissian}}}, \ and\ \bibinfo
  {author} {\bibfnamefont {F.}~\bibnamefont {{Camilo}}},\ }\href {\doibase
  10.1086/382155} {\bibfield  {journal} {\bibinfo  {journal} {\apjl}\ }\textbf
  {\bibinfo {volume} {601}},\ \bibinfo {pages} {L179} (\bibinfo {year}
  {2004}{\natexlab{a}})}%
\bibitem [{\citenamefont {{Kalogera}}\ \emph
  {et~al.}(2004{\natexlab{b}})\citenamefont {{Kalogera}}, \citenamefont
  {{Kim}}, \citenamefont {{Lorimer}}, \citenamefont {{Burgay}}, \citenamefont
  {{D'Amico}}, \citenamefont {{Possenti}}, \citenamefont {{Manchester}},
  \citenamefont {{Lyne}}, \citenamefont {{Joshi}}, \citenamefont
  {{McLaughlin}}, \citenamefont {{Kramer}}, \citenamefont {{Sarkissian}},\ and\
  \citenamefont {{Camilo}}}]{KalEtal04a}%
  \BibitemOpen
  \bibfield  {author} {\bibinfo {author} {\bibfnamefont {V.}~\bibnamefont
  {{Kalogera}}}, \bibinfo {author} {\bibfnamefont {C.}~\bibnamefont {{Kim}}},
  \bibinfo {author} {\bibfnamefont {D.~R.}\ \bibnamefont {{Lorimer}}}, \bibinfo
  {author} {\bibfnamefont {M.}~\bibnamefont {{Burgay}}}, \bibinfo {author}
  {\bibfnamefont {N.}~\bibnamefont {{D'Amico}}}, \bibinfo {author}
  {\bibfnamefont {A.}~\bibnamefont {{Possenti}}}, \bibinfo {author}
  {\bibfnamefont {R.~N.}\ \bibnamefont {{Manchester}}}, \bibinfo {author}
  {\bibfnamefont {A.~G.}\ \bibnamefont {{Lyne}}}, \bibinfo {author}
  {\bibfnamefont {B.~C.}\ \bibnamefont {{Joshi}}}, \bibinfo {author}
  {\bibfnamefont {M.~A.}\ \bibnamefont {{McLaughlin}}}, \bibinfo {author}
  {\bibfnamefont {M.}~\bibnamefont {{Kramer}}}, \bibinfo {author}
  {\bibfnamefont {J.~M.}\ \bibnamefont {{Sarkissian}}}, \ and\ \bibinfo
  {author} {\bibfnamefont {F.}~\bibnamefont {{Camilo}}},\ }\href {\doibase
  10.1086/425868} {\bibfield  {journal} {\bibinfo  {journal} {\apjl}\ }\textbf
  {\bibinfo {volume} {614}},\ \bibinfo {pages} {L137} (\bibinfo {year}
  {2004}{\natexlab{b}})},\ \Eprint
  {http://arxiv.org/abs/arXiv:astro-ph/0312101} {arXiv:astro-ph/0312101}
  %
\bibitem [{\citenamefont {{Abadie}}\ \emph {et~al.}(2010)\citenamefont
  {{Abadie}}, \citenamefont {{Abbott}}, \citenamefont {{Abbott}}, \citenamefont
  {{Abernathy}}, \citenamefont {{Accadia}}, \citenamefont {{Acernese}},
  \citenamefont {{Adams}}, \citenamefont {{Adhikari}}, \citenamefont {{Ajith}},
  \citenamefont {{Allen}},\ and\ \citenamefont {et~al.}}]{abadie10}%
  \BibitemOpen
  \bibfield  {author} {\bibinfo {author} {\bibfnamefont {J.}~\bibnamefont
  {{Abadie}}}, \bibinfo {author} {\bibfnamefont {B.~P.}\ \bibnamefont
  {{Abbott}}}, \bibinfo {author} {\bibfnamefont {R.}~\bibnamefont {{Abbott}}},
  \bibinfo {author} {\bibfnamefont {M.}~\bibnamefont {{Abernathy}}}, \bibinfo
  {author} {\bibfnamefont {T.}~\bibnamefont {{Accadia}}}, \bibinfo {author}
  {\bibfnamefont {F.}~\bibnamefont {{Acernese}}}, \bibinfo {author}
  {\bibfnamefont {C.}~\bibnamefont {{Adams}}}, \bibinfo {author} {\bibfnamefont
  {R.}~\bibnamefont {{Adhikari}}}, \bibinfo {author} {\bibfnamefont
  {P.}~\bibnamefont {{Ajith}}}, \bibinfo {author} {\bibfnamefont
  {B.}~\bibnamefont {{Allen}}}, \ and\ \bibinfo {author} {\bibnamefont
  {et~al.}},\ }\href {\doibase 10.1088/0264-9381/27/17/173001} {\bibfield
  {journal} {\bibinfo  {journal} {Classical and Quantum Gravity}\ }\textbf
  {\bibinfo {volume} {27}},\ \bibinfo {pages} {173001} (\bibinfo {year}
  {2010})},\ \Eprint {http://arxiv.org/abs/1003.2480} {arXiv:1003.2480
  [astro-ph.HE]} %
\bibitem [{\citenamefont {{Gratton}}\ and\ \citenamefont
  {{Sneden}}(1994)}]{gratton94}%
  \BibitemOpen
  \bibfield  {author} {\bibinfo {author} {\bibfnamefont {R.~G.}\ \bibnamefont
  {{Gratton}}}\ and\ \bibinfo {author} {\bibfnamefont {C.}~\bibnamefont
  {{Sneden}}},\ }\href@noop {} {\bibfield  {journal} {\bibinfo  {journal}
  {\aap}\ }\textbf {\bibinfo {volume} {287}},\ \bibinfo {pages} {927} (\bibinfo
  {year} {1994})}%
\bibitem [{\citenamefont {{Crawford}}\ \emph {et~al.}(1998)\citenamefont
  {{Crawford}}, \citenamefont {{Sneden}}, \citenamefont {{King}}, \citenamefont
  {{Boesgaard}},\ and\ \citenamefont {{Deliyannis}}}]{crawford98}%
  \BibitemOpen
  \bibfield  {author} {\bibinfo {author} {\bibfnamefont {J.~L.}\ \bibnamefont
  {{Crawford}}}, \bibinfo {author} {\bibfnamefont {C.}~\bibnamefont
  {{Sneden}}}, \bibinfo {author} {\bibfnamefont {J.~R.}\ \bibnamefont
  {{King}}}, \bibinfo {author} {\bibfnamefont {A.~M.}\ \bibnamefont
  {{Boesgaard}}}, \ and\ \bibinfo {author} {\bibfnamefont {C.~P.}\ \bibnamefont
  {{Deliyannis}}},\ }\href {\doibase 10.1086/300600} {\bibfield  {journal}
  {\bibinfo  {journal} {\aj}\ }\textbf {\bibinfo {volume} {116}},\ \bibinfo
  {pages} {2489} (\bibinfo {year} {1998})}%
\bibitem [{\citenamefont {{McWilliam}}\ \emph {et~al.}(1995)\citenamefont
  {{McWilliam}}, \citenamefont {{Preston}}, \citenamefont {{Sneden}},\ and\
  \citenamefont {{Shectman}}}]{mcwilliam95a}%
  \BibitemOpen
  \bibfield  {author} {\bibinfo {author} {\bibfnamefont {A.}~\bibnamefont
  {{McWilliam}}}, \bibinfo {author} {\bibfnamefont {G.~W.}\ \bibnamefont
  {{Preston}}}, \bibinfo {author} {\bibfnamefont {C.}~\bibnamefont {{Sneden}}},
  \ and\ \bibinfo {author} {\bibfnamefont {S.}~\bibnamefont {{Shectman}}},\
  }\href {\doibase 10.1086/117485} {\bibfield  {journal} {\bibinfo  {journal}
  {\aj}\ }\textbf {\bibinfo {volume} {109}},\ \bibinfo {pages} {2736} (\bibinfo
  {year} {1995})}%
\bibitem [{\citenamefont {{Cresci}}\ \emph {et~al.}(2009)\citenamefont
  {{Cresci}}, \citenamefont {{Hicks}}, \citenamefont {{Genzel}}, \citenamefont
  {{Schreiber}}, \citenamefont {{Davies}}, \citenamefont {{Bouch{\'e}}},
  \citenamefont {{Buschkamp}}, \citenamefont {{Genel}}, \citenamefont
  {{Shapiro}}, \citenamefont {{Tacconi}}, \citenamefont {{Sommer-Larsen}},
  \citenamefont {{Shapley}}, \citenamefont {{Steidel}},\ and\ \citenamefont
  {{Caputi}}}]{cresci09}%
  \BibitemOpen
  \bibfield  {author} {\bibinfo {author} {\bibfnamefont {G.}~\bibnamefont
  {{Cresci}}}, \bibinfo {author} {\bibfnamefont {E.~K.~S.}\ \bibnamefont
  {{Hicks}}}, \bibinfo {author} {\bibfnamefont {R.}~\bibnamefont {{Genzel}}},
  \bibinfo {author} {\bibfnamefont {N.~M.~F.}\ \bibnamefont {{Schreiber}}},
  \bibinfo {author} {\bibfnamefont {R.}~\bibnamefont {{Davies}}}, \bibinfo
  {author} {\bibfnamefont {N.}~\bibnamefont {{Bouch{\'e}}}}, \bibinfo {author}
  {\bibfnamefont {P.}~\bibnamefont {{Buschkamp}}}, \bibinfo {author}
  {\bibfnamefont {S.}~\bibnamefont {{Genel}}}, \bibinfo {author} {\bibfnamefont
  {K.}~\bibnamefont {{Shapiro}}}, \bibinfo {author} {\bibfnamefont
  {L.}~\bibnamefont {{Tacconi}}}, \bibinfo {author} {\bibfnamefont
  {J.}~\bibnamefont {{Sommer-Larsen}}}, \bibinfo {author} {\bibfnamefont
  {A.}~\bibnamefont {{Shapley}}}, \bibinfo {author} {\bibfnamefont {C.~C.}\
  \bibnamefont {{Steidel}}}, \ and\ \bibinfo {author} {\bibfnamefont
  {K.}~\bibnamefont {{Caputi}}},\ }\href {\doibase 10.1088/0004-637X/697/1/115}
  {\bibfield  {journal} {\bibinfo  {journal} {\apj}\ }\textbf {\bibinfo
  {volume} {697}},\ \bibinfo {pages} {115} (\bibinfo {year} {2009})},\ \Eprint
  {http://arxiv.org/abs/0902.4701} {arXiv:0902.4701 [astro-ph.CO]} 
  %
\bibitem [{\citenamefont {{Narayan}}\ \emph {et~al.}(1992)\citenamefont
  {{Narayan}}, \citenamefont {{Paczynski}},\ and\ \citenamefont
  {{Piran}}}]{narayan92}%
  \BibitemOpen
  \bibfield  {author} {\bibinfo {author} {\bibfnamefont {R.}~\bibnamefont
  {{Narayan}}}, \bibinfo {author} {\bibfnamefont {B.}~\bibnamefont
  {{Paczynski}}}, \ and\ \bibinfo {author} {\bibfnamefont {T.}~\bibnamefont
  {{Piran}}},\ }\href {\doibase 10.1086/186493} {\bibfield  {journal} {\bibinfo
   {journal} {\apjl}\ }\textbf {\bibinfo {volume} {395}},\ \bibinfo {pages}
  {L83} (\bibinfo {year} {1992})},\ \Eprint
  {http://arxiv.org/abs/astro-ph/9204001} {astro-ph/9204001} 
  %
\bibitem [{\citenamefont {{Fryer}}\ \emph {et~al.}(1999)\citenamefont
  {{Fryer}}, \citenamefont {{Woosley}},\ and\ \citenamefont
  {{Hartmann}}}]{fryer99a}%
  \BibitemOpen
  \bibfield  {author} {\bibinfo {author} {\bibfnamefont {C.~L.}\ \bibnamefont
  {{Fryer}}}, \bibinfo {author} {\bibfnamefont {S.~E.}\ \bibnamefont
  {{Woosley}}}, \ and\ \bibinfo {author} {\bibfnamefont {D.~H.}\ \bibnamefont
  {{Hartmann}}},\ }\href@noop {} {\bibfield  {journal} {\bibinfo  {journal}
  {ApJ}\ }\textbf {\bibinfo {volume} {526}},\ \bibinfo {pages} {152} (\bibinfo
  {year} {1999})}%
\bibitem [{\citenamefont {{Bloom}}\ \emph {et~al.}(2002)\citenamefont
  {{Bloom}}, \citenamefont {{Kulkarni}},\ and\ \citenamefont
  {{Djorgovski}}}]{bloom02}%
  \BibitemOpen
  \bibfield  {author} {\bibinfo {author} {\bibfnamefont {J.~S.}\ \bibnamefont
  {{Bloom}}}, \bibinfo {author} {\bibfnamefont {S.~R.}\ \bibnamefont
  {{Kulkarni}}}, \ and\ \bibinfo {author} {\bibfnamefont {S.~G.}\ \bibnamefont
  {{Djorgovski}}},\ }\href {\doibase 10.1086/338893} {\bibfield  {journal}
  {\bibinfo  {journal} {\aj}\ }\textbf {\bibinfo {volume} {123}},\ \bibinfo
  {pages} {1111} (\bibinfo {year} {2002})},\ \Eprint
  {http://arxiv.org/abs/astro-ph/0010176} {astro-ph/0010176} 
  %
\bibitem [{\citenamefont {{Rosswog}}\ \emph {et~al.}(2003)\citenamefont
  {{Rosswog}}, \citenamefont {{Ramirez-Ruiz}},\ and\ \citenamefont
  {{Davies}}}]{rosswog03c}%
  \BibitemOpen
  \bibfield  {author} {\bibinfo {author} {\bibfnamefont {S.}~\bibnamefont
  {{Rosswog}}}, \bibinfo {author} {\bibfnamefont {E.}~\bibnamefont
  {{Ramirez-Ruiz}}}, \ and\ \bibinfo {author} {\bibfnamefont {M.~B.}\
  \bibnamefont {{Davies}}},\ }\href {\doibase 10.1046/j.1365-2966.2003.07032.x}
  {\bibfield  {journal} {\bibinfo  {journal} {MNRAS}\ }\textbf {\bibinfo
  {volume} {345}},\ \bibinfo {pages} {1077} (\bibinfo {year}
  {2003})}%
\bibitem [{\citenamefont {{Fong}}\ \emph {et~al.}(2010)\citenamefont {{Fong}},
  \citenamefont {{Berger}},\ and\ \citenamefont {{Fox}}}]{fong10}%
  \BibitemOpen
  \bibfield  {author} {\bibinfo {author} {\bibfnamefont {W.}~\bibnamefont
  {{Fong}}}, \bibinfo {author} {\bibfnamefont {E.}~\bibnamefont {{Berger}}}, \
  and\ \bibinfo {author} {\bibfnamefont {D.~B.}\ \bibnamefont {{Fox}}},\ }\href
  {\doibase 10.1088/0004-637X/708/1/9} {\bibfield  {journal} {\bibinfo
  {journal} {ApJ}\ }\textbf {\bibinfo {volume} {708}},\ \bibinfo {pages} {9}
  (\bibinfo {year} {2010})},\ \Eprint {http://arxiv.org/abs/0909.1804}
  {arXiv:0909.1804 [astro-ph.HE]} %
\bibitem [{\citenamefont {{Woolf}}\ \emph {et~al.}(1995)\citenamefont
  {{Woolf}}, \citenamefont {{Tomkin}},\ and\ \citenamefont
  {{Lambert}}}]{woolf95}%
  \BibitemOpen
  \bibfield  {author} {\bibinfo {author} {\bibfnamefont {V.~M.}\ \bibnamefont
  {{Woolf}}}, \bibinfo {author} {\bibfnamefont {J.}~\bibnamefont {{Tomkin}}}, \
  and\ \bibinfo {author} {\bibfnamefont {D.~L.}\ \bibnamefont {{Lambert}}},\
  }\href {\doibase 10.1086/176428} {\bibfield  {journal} {\bibinfo  {journal}
  {\apj}\ }\textbf {\bibinfo {volume} {453}},\ \bibinfo {pages} {660} (\bibinfo
  {year} {1995})}%
\bibitem [{\citenamefont {{Shetrone}}(1996)}]{Shetrone96}%
  \BibitemOpen
  \bibfield  {author} {\bibinfo {author} {\bibfnamefont {M.~D.}\ \bibnamefont
  {{Shetrone}}},\ }\href {\doibase 10.1086/118120} {\bibfield  {journal}
  {\bibinfo  {journal} {\aj}\ }\textbf {\bibinfo {volume} {112}},\ \bibinfo
  {pages} {1517} (\bibinfo {year} {1996})}%
\bibitem [{\citenamefont {{Sneden}}\ \emph {et~al.}(2000)\citenamefont
  {{Sneden}}, \citenamefont {{Cowan}}, \citenamefont {{Beers}}, \citenamefont
  {{Burles}}, \citenamefont {{Fuller}}, \citenamefont {{Ivans}},\ and\
  \citenamefont {{Lawler}}}]{sneden00}%
  \BibitemOpen
  \bibfield  {author} {\bibinfo {author} {\bibfnamefont {C.}~\bibnamefont
  {{Sneden}}}, \bibinfo {author} {\bibfnamefont {J.~J.}\ \bibnamefont
  {{Cowan}}}, \bibinfo {author} {\bibfnamefont {T.~C.}\ \bibnamefont
  {{Beers}}}, \bibinfo {author} {\bibfnamefont {S.}~\bibnamefont {{Burles}}},
  \bibinfo {author} {\bibfnamefont {G.~M.}\ \bibnamefont {{Fuller}}}, \bibinfo
  {author} {\bibfnamefont {I.~I.}\ \bibnamefont {{Ivans}}}, \ and\ \bibinfo
  {author} {\bibfnamefont {J.~E.}\ \bibnamefont {{Lawler}}},\ }in\ \href@noop
  {}  \bibinfo {series} {Bulletin of the American Astronomical
  Society}, Vol.~\bibinfo {volume} {32}\ (\bibinfo {year} {2000})\ p.\ \bibinfo
  {pages} {1475}%
\bibitem [{\citenamefont {{Burris}}\ \emph {et~al.}(2000)\citenamefont
  {{Burris}}, \citenamefont {{Pilachowski}}, \citenamefont {{Armandroff}},
  \citenamefont {{Sneden}}, \citenamefont {{Cowan}},\ and\ \citenamefont
  {{Roe}}}]{Burris00}%
  \BibitemOpen
  \bibfield  {author} {\bibinfo {author} {\bibfnamefont {D.~L.}\ \bibnamefont
  {{Burris}}}, \bibinfo {author} {\bibfnamefont {C.~A.}\ \bibnamefont
  {{Pilachowski}}}, \bibinfo {author} {\bibfnamefont {T.~E.}\ \bibnamefont
  {{Armandroff}}}, \bibinfo {author} {\bibfnamefont {C.}~\bibnamefont
  {{Sneden}}}, \bibinfo {author} {\bibfnamefont {J.~J.}\ \bibnamefont
  {{Cowan}}}, \ and\ \bibinfo {author} {\bibfnamefont {H.}~\bibnamefont
  {{Roe}}},\ }\href {\doibase 10.1086/317172} {\bibfield  {journal} {\bibinfo
  {journal} {\apj}\ }\textbf {\bibinfo {volume} {544}},\ \bibinfo {pages} {302}
  (\bibinfo {year} {2000})},\ \Eprint {http://arxiv.org/abs/astro-ph/0005188}
  {astro-ph/0005188} %
\bibitem [{\citenamefont {{Cayrel}}\ \emph {et~al.}(2001)\citenamefont
  {{Cayrel}}, \citenamefont {{Spite}}, \citenamefont {{Spite}}, \citenamefont
  {{Hill}}, \citenamefont {{Primas}}, \citenamefont {{Fran{\c c}ois}},
  \citenamefont {{Beers}}, \citenamefont {{Plez}}, \citenamefont {{Barbuy}},
  \citenamefont {{Andersen}}, \citenamefont {{Nordstr{\"o}m}}, \citenamefont
  {{Molaro}},\ and\ \citenamefont {{Bonifacio}}}]{Cayrel01}%
  \BibitemOpen
  \bibfield  {author} {\bibinfo {author} {\bibfnamefont {R.}~\bibnamefont
  {{Cayrel}}}, \bibinfo {author} {\bibfnamefont {M.}~\bibnamefont {{Spite}}},
  \bibinfo {author} {\bibfnamefont {F.}~\bibnamefont {{Spite}}}, \bibinfo
  {author} {\bibfnamefont {V.}~\bibnamefont {{Hill}}}, \bibinfo {author}
  {\bibfnamefont {F.}~\bibnamefont {{Primas}}}, \bibinfo {author}
  {\bibfnamefont {P.}~\bibnamefont {{Fran{\c c}ois}}}, \bibinfo {author}
  {\bibfnamefont {T.~C.}\ \bibnamefont {{Beers}}}, \bibinfo {author}
  {\bibfnamefont {B.}~\bibnamefont {{Plez}}}, \bibinfo {author} {\bibfnamefont
  {B.}~\bibnamefont {{Barbuy}}}, \bibinfo {author} {\bibfnamefont
  {J.}~\bibnamefont {{Andersen}}}, \bibinfo {author} {\bibfnamefont
  {B.}~\bibnamefont {{Nordstr{\"o}m}}}, \bibinfo {author} {\bibfnamefont
  {P.}~\bibnamefont {{Molaro}}}, \ and\ \bibinfo {author} {\bibfnamefont
  {P.}~\bibnamefont {{Bonifacio}}},\ }in\ \href@noop {} { {\bibinfo
  {booktitle} {Astrophysical Ages and Times Scales}}},\ \bibinfo {series}
  {Astronomical Society of the Pacific Conference Series}, Vol.\ \bibinfo
  {volume} {245},\ \bibinfo {editor} {edited by\ \bibinfo {editor}
  {\bibfnamefont {T.}~\bibnamefont {{von Hippel}}}, \bibinfo {editor}
  {\bibfnamefont {C.}~\bibnamefont {{Simpson}}}, \ and\ \bibinfo {editor}
  {\bibfnamefont {N.}~\bibnamefont {{Manset}}}}\ (\bibinfo {year} {2001})\ p.\
  \bibinfo {pages} {244},\ \Eprint {http://arxiv.org/abs/astro-ph/0104448}
  {astro-ph/0104448} %
\bibitem [{\citenamefont {{Hill}}\ \emph {et~al.}(2002)\citenamefont {{Hill}},
  \citenamefont {{Plez}}, \citenamefont {{Cayrel}}, \citenamefont {{Beers}},
  \citenamefont {{Nordstr{\"o}m}}, \citenamefont {{Andersen}}, \citenamefont
  {{Spite}}, \citenamefont {{Spite}}, \citenamefont {{Barbuy}}, \citenamefont
  {{Bonifacio}}, \citenamefont {{Depagne}}, \citenamefont {{Fran{\c c}ois}},\
  and\ \citenamefont {{Primas}}}]{Hill02}%
  \BibitemOpen
  \bibfield  {author} {\bibinfo {author} {\bibfnamefont {V.}~\bibnamefont
  {{Hill}}}, \bibinfo {author} {\bibfnamefont {B.}~\bibnamefont {{Plez}}},
  \bibinfo {author} {\bibfnamefont {R.}~\bibnamefont {{Cayrel}}}, \bibinfo
  {author} {\bibfnamefont {T.~C.}\ \bibnamefont {{Beers}}}, \bibinfo {author}
  {\bibfnamefont {B.}~\bibnamefont {{Nordstr{\"o}m}}}, \bibinfo {author}
  {\bibfnamefont {J.}~\bibnamefont {{Andersen}}}, \bibinfo {author}
  {\bibfnamefont {M.}~\bibnamefont {{Spite}}}, \bibinfo {author} {\bibfnamefont
  {F.}~\bibnamefont {{Spite}}}, \bibinfo {author} {\bibfnamefont
  {B.}~\bibnamefont {{Barbuy}}}, \bibinfo {author} {\bibfnamefont
  {P.}~\bibnamefont {{Bonifacio}}}, \bibinfo {author} {\bibfnamefont
  {E.}~\bibnamefont {{Depagne}}}, \bibinfo {author} {\bibfnamefont
  {P.}~\bibnamefont {{Fran{\c c}ois}}}, \ and\ \bibinfo {author} {\bibfnamefont
  {F.}~\bibnamefont {{Primas}}},\ }\href {\doibase 10.1051/0004-6361:20020434}
  {\bibfield  {journal} {\bibinfo  {journal} {\aap}\ }\textbf {\bibinfo
  {volume} {387}},\ \bibinfo {pages} {560} (\bibinfo {year} {2002})},\ \Eprint
  {http://arxiv.org/abs/astro-ph/0203462} {astro-ph/0203462} 
  %
\bibitem [{\citenamefont {{Belczynski}}\ \emph {et~al.}(2002)\citenamefont
  {{Belczynski}}, \citenamefont {{Kalogera}},\ and\ \citenamefont
  {{Bulik}}}]{belczynski02b}%
  \BibitemOpen
  \bibfield  {author} {\bibinfo {author} {\bibfnamefont {K.}~\bibnamefont
  {{Belczynski}}}, \bibinfo {author} {\bibfnamefont {V.}~\bibnamefont
  {{Kalogera}}}, \ and\ \bibinfo {author} {\bibfnamefont {T.}~\bibnamefont
  {{Bulik}}},\ }\href {\doibase 10.1086/340304} {\bibfield  {journal} {\bibinfo
   {journal} {\apj}\ }\textbf {\bibinfo {volume} {572}},\ \bibinfo {pages}
  {407} (\bibinfo {year} {2002})},\ \Eprint
  {http://arxiv.org/abs/astro-ph/0111452} {astro-ph/0111452} 
  %
\bibitem [{\citenamefont {{Wasserburg}}\ \emph {et~al.}(2006)\citenamefont
  {{Wasserburg}}, \citenamefont {{Busso}}, \citenamefont {{Gallino}},\ and\
  \citenamefont {{Nollett}}}]{wasserburg06}%
  \BibitemOpen
  \bibfield  {author} {\bibinfo {author} {\bibfnamefont {G.~J.}\ \bibnamefont
  {{Wasserburg}}}, \bibinfo {author} {\bibfnamefont {M.}~\bibnamefont
  {{Busso}}}, \bibinfo {author} {\bibfnamefont {R.}~\bibnamefont {{Gallino}}},
  \ and\ \bibinfo {author} {\bibfnamefont {K.~M.}\ \bibnamefont {{Nollett}}},\
  }\href {\doibase 10.1016/j.nuclphysa.2005.07.015} {\bibfield  {journal}
  {\bibinfo  {journal} {Nuclear Physics A}\ }\textbf {\bibinfo {volume}
  {777}},\ \bibinfo {pages} {5} (\bibinfo {year} {2006})},\ \Eprint
  {http://arxiv.org/abs/astro-ph/0602551} {astro-ph/0602551} 
  %
\bibitem [{\citenamefont {{Paul}}\ \emph {et~al.}(2001)\citenamefont {{Paul}},
  \citenamefont {{Valenta}}, \citenamefont {{Ahmad}}, \citenamefont
  {{Berkovits}}, \citenamefont {{Bordeanu}}, \citenamefont {{Ghelberg}},
  \citenamefont {{Hashimoto}}, \citenamefont {{Hershkowitz}}, \citenamefont
  {{Jiang}}, \citenamefont {{Nakanishi}},\ and\ \citenamefont
  {{Sakamoto}}}]{paul01}%
  \BibitemOpen
  \bibfield  {author} {\bibinfo {author} {\bibfnamefont {M.}~\bibnamefont
  {{Paul}}}, \bibinfo {author} {\bibfnamefont {A.}~\bibnamefont {{Valenta}}},
  \bibinfo {author} {\bibfnamefont {I.}~\bibnamefont {{Ahmad}}}, \bibinfo
  {author} {\bibfnamefont {D.}~\bibnamefont {{Berkovits}}}, \bibinfo {author}
  {\bibfnamefont {C.}~\bibnamefont {{Bordeanu}}}, \bibinfo {author}
  {\bibfnamefont {S.}~\bibnamefont {{Ghelberg}}}, \bibinfo {author}
  {\bibfnamefont {Y.}~\bibnamefont {{Hashimoto}}}, \bibinfo {author}
  {\bibfnamefont {A.}~\bibnamefont {{Hershkowitz}}}, \bibinfo {author}
  {\bibfnamefont {S.}~\bibnamefont {{Jiang}}}, \bibinfo {author} {\bibfnamefont
  {T.}~\bibnamefont {{Nakanishi}}}, \ and\ \bibinfo {author} {\bibfnamefont
  {K.}~\bibnamefont {{Sakamoto}}},\ }\href {\doibase 10.1086/323617} {\bibfield
   {journal} {\bibinfo  {journal} {\apjl}\ }\textbf {\bibinfo {volume} {558}},\
  \bibinfo {pages} {L133} (\bibinfo {year} {2001})},\ \Eprint
  {http://arxiv.org/abs/astro-ph/0106205} {astro-ph/0106205} 
  %
\bibitem [{\citenamefont {{Wallner}}\ \emph {et~al.}(2014)\citenamefont
  {{Wallner}}, \citenamefont {{Faestermann}}, \citenamefont {{Feldstein}},
  \citenamefont {{Knie}}, \citenamefont {{Korschinek}}, \citenamefont
  {{Kutschera}}, \citenamefont {{Ofan}}, \citenamefont {{Paul}}, \citenamefont
  {{Quinto}}, \citenamefont {{Rugel}},\ and\ \citenamefont
  {{Steier}}}]{walner14}%
  \BibitemOpen
  \bibfield  {author} {\bibinfo {author} {\bibfnamefont {A.}~\bibnamefont
  {{Wallner}}}, \bibinfo {author} {\bibfnamefont {T.}~\bibnamefont
  {{Faestermann}}}, \bibinfo {author} {\bibfnamefont {C.}~\bibnamefont
  {{Feldstein}}}, \bibinfo {author} {\bibfnamefont {K.}~\bibnamefont {{Knie}}},
  \bibinfo {author} {\bibfnamefont {G.}~\bibnamefont {{Korschinek}}}, \bibinfo
  {author} {\bibfnamefont {W.}~\bibnamefont {{Kutschera}}}, \bibinfo {author}
  {\bibfnamefont {A.}~\bibnamefont {{Ofan}}}, \bibinfo {author} {\bibfnamefont
  {M.}~\bibnamefont {{Paul}}}, \bibinfo {author} {\bibfnamefont
  {F.}~\bibnamefont {{Quinto}}}, \bibinfo {author} {\bibfnamefont
  {G.}~\bibnamefont {{Rugel}}}, \ and\ \bibinfo {author} {\bibfnamefont
  {P.}~\bibnamefont {{Steier}}},\ }\href@noop {} {\bibfield  {journal}
  {\bibinfo  {journal} {In preparation}\ } (\bibinfo {year}
  {2014})}%
\bibitem [{\citenamefont {{Hotokezaka}}\ \emph {et~al.}(2013)\citenamefont
  {{Hotokezaka}}, \citenamefont {{Kiuchi}}, \citenamefont {{Kyutoku}},
  \citenamefont {{Okawa}}, \citenamefont {{Sekiguchi}}, \citenamefont
  {{Shibata}},\ and\ \citenamefont {{Taniguchi}}}]{hotokezaka13b}%
  \BibitemOpen
  \bibfield  {author} {\bibinfo {author} {\bibfnamefont {K.}~\bibnamefont
  {{Hotokezaka}}}, \bibinfo {author} {\bibfnamefont {K.}~\bibnamefont
  {{Kiuchi}}}, \bibinfo {author} {\bibfnamefont {K.}~\bibnamefont {{Kyutoku}}},
  \bibinfo {author} {\bibfnamefont {H.}~\bibnamefont {{Okawa}}}, \bibinfo
  {author} {\bibfnamefont {Y.-i.}\ \bibnamefont {{Sekiguchi}}}, \bibinfo
  {author} {\bibfnamefont {M.}~\bibnamefont {{Shibata}}}, \ and\ \bibinfo
  {author} {\bibfnamefont {K.}~\bibnamefont {{Taniguchi}}},\ }\href {\doibase
  10.1103/PhysRevD.87.024001} {\bibfield  {journal} {\bibinfo  {journal} {Phys.
  Rev. D}\ }\textbf {\bibinfo {volume} {87}},\ \bibinfo {eid} {024001}
  (\bibinfo {year} {2013})},\ \Eprint {http://arxiv.org/abs/1212.0905}
  {arXiv:1212.0905 [astro-ph.HE]} %
\bibitem [{\citenamefont {{Lovelace}}\ \emph {et~al.}(2013)\citenamefont
  {{Lovelace}}, \citenamefont {{Duez}}, \citenamefont {{Foucart}},
  \citenamefont {{Kidder}}, \citenamefont {{Pfeiffer}}, \citenamefont
  {{Scheel}},\ and\ \citenamefont {{Szil{\'a}gyi}}}]{lovelace13}%
  \BibitemOpen
  \bibfield  {author} {\bibinfo {author} {\bibfnamefont {G.}~\bibnamefont
  {{Lovelace}}}, \bibinfo {author} {\bibfnamefont {M.~D.}\ \bibnamefont
  {{Duez}}}, \bibinfo {author} {\bibfnamefont {F.}~\bibnamefont {{Foucart}}},
  \bibinfo {author} {\bibfnamefont {L.~E.}\ \bibnamefont {{Kidder}}}, \bibinfo
  {author} {\bibfnamefont {H.~P.}\ \bibnamefont {{Pfeiffer}}}, \bibinfo
  {author} {\bibfnamefont {M.~A.}\ \bibnamefont {{Scheel}}}, \ and\ \bibinfo
  {author} {\bibfnamefont {B.}~\bibnamefont {{Szil{\'a}gyi}}},\ }\href
  {\doibase 10.1088/0264-9381/30/13/135004} {\bibfield  {journal} {\bibinfo
  {journal} {Classical and Quantum Gravity}\ }\textbf {\bibinfo {volume}
  {30}},\ \bibinfo {eid} {135004} (\bibinfo {year} {2013})},\ \Eprint
  {http://arxiv.org/abs/1302.6297} {arXiv:1302.6297 [gr-qc]} 
  %
\bibitem [{\citenamefont {{Deaton}}\ \emph {et~al.}(2013)\citenamefont
  {{Deaton}}, \citenamefont {{Duez}}, \citenamefont {{Foucart}}, \citenamefont
  {{O'Connor}}, \citenamefont {{Ott}}, \citenamefont {{Kidder}}, \citenamefont
  {{Muhlberger}}, \citenamefont {{Scheel}},\ and\ \citenamefont
  {{Szilagyi}}}]{deaton13}%
  \BibitemOpen
  \bibfield  {author} {\bibinfo {author} {\bibfnamefont {M.~B.}\ \bibnamefont
  {{Deaton}}}, \bibinfo {author} {\bibfnamefont {M.~D.}\ \bibnamefont
  {{Duez}}}, \bibinfo {author} {\bibfnamefont {F.}~\bibnamefont {{Foucart}}},
  \bibinfo {author} {\bibfnamefont {E.}~\bibnamefont {{O'Connor}}}, \bibinfo
  {author} {\bibfnamefont {C.~D.}\ \bibnamefont {{Ott}}}, \bibinfo {author}
  {\bibfnamefont {L.~E.}\ \bibnamefont {{Kidder}}}, \bibinfo {author}
  {\bibfnamefont {C.~D.}\ \bibnamefont {{Muhlberger}}}, \bibinfo {author}
  {\bibfnamefont {M.~A.}\ \bibnamefont {{Scheel}}}, \ and\ \bibinfo {author}
  {\bibfnamefont {B.}~\bibnamefont {{Szilagyi}}},\ }\href {\doibase
  10.1088/0004-637X/776/1/47} {\bibfield  {journal} {\bibinfo  {journal}
  {\apj}\ }\textbf {\bibinfo {volume} {776}},\ \bibinfo {eid} {47} (\bibinfo
  {year} {2013})},\ \Eprint {http://arxiv.org/abs/1304.3384} {arXiv:1304.3384
  [astro-ph.HE]} %
\bibitem [{\citenamefont {{Lee}}\ \emph {et~al.}(2010)\citenamefont {{Lee}},
  \citenamefont {{Ramirez-Ruiz}},\ and\ \citenamefont {{van de Ven}}}]{lee10a}%
  \BibitemOpen
  \bibfield  {author} {\bibinfo {author} {\bibfnamefont {W.~H.}\ \bibnamefont
  {{Lee}}}, \bibinfo {author} {\bibfnamefont {E.}~\bibnamefont
  {{Ramirez-Ruiz}}}, \ and\ \bibinfo {author} {\bibfnamefont {G.}~\bibnamefont
  {{van de Ven}}},\ }\href {\doibase 10.1088/0004-637X/720/1/953} {\bibfield
  {journal} {\bibinfo  {journal} {ApJ}\ }\textbf {\bibinfo {volume} {720}},\
  \bibinfo {pages} {953} (\bibinfo {year} {2010})},\ \Eprint
  {http://arxiv.org/abs/0909.2884} {arXiv:0909.2884 [astro-ph.HE]} 
  %
\bibitem [{\citenamefont {{Nakar}}\ and\ \citenamefont {{Piran}}(2011)}]{NP11}%
  \BibitemOpen
  \bibfield  {author} {\bibinfo {author} {\bibfnamefont {E.}~\bibnamefont
  {{Nakar}}}\ and\ \bibinfo {author} {\bibfnamefont {T.}~\bibnamefont
  {{Piran}}},\ }\href {\doibase 10.1038/nature10365} {\bibfield  {journal}
  {\bibinfo  {journal} {\nat}\ }\textbf {\bibinfo {volume} {478}},\ \bibinfo
  {pages} {82} (\bibinfo {year} {2011})}%
\end{thebibliography}

\hyphenation{Post-Script Sprin-ger}
\end{document}